\documentclass{aa}
\usepackage{natbib}
\usepackage{txfonts}
\usepackage{graphicx}

\begin{document}

   \title{Hybrid morphology radio sources from the FIRST survey}

   \author{M.~P. Gawro\'nski\inst{1}
   \and A. Marecki\inst{1}
   \and M. Kunert-Bajraszewska\inst{1}
   \and A.~J. Kus\inst{1}} 

   \offprints{A. Marecki\\ \email{amr@astro.uni.torun.pl}}
   
   \institute{Toru\'n Centre for Astronomy, N. Copernicus University,
             87-100 Toru\'n, Poland}

   \date{Received 5 August 2005 / Accepted 13 September 2005}

\abstract{The so-called HYbrid MOrphology Radio Sources (HYMORS) are a class of
objects that appear to have a mixed Fanaroff-Riley (FR) morphology in a single
object; i.e. a HYMORS has an FR\,I-type lobe on one side of its nucleus and
an FR\,II-type lobe on the other side. Because of this unique feature and given
that the origin of the FR morphological dichotomy is still unclear, HYMORS may
possibly play a crucial role in our understanding of the FR-dichotomy. As the
number of known HYMORS is quite small, we aimed to increase that number by
inspecting a few areas of the sky covered by the VLA FIRST survey and by
selecting 21~HYMORS candidates based on the morphology shown in the FIRST
images. They were observed with the VLA in B-conf. at
4.9\,GHz. Three objects from the initial sample turned out to be actual 
HYMORS and two others very likely to fulfill the criteria. These five were 
subsequently re-observed with the VLA in A-conf. at 1.4\,GHz. Our results 
provide strong support to the findings of \citet{gkw2000}, namely that there 
are two different kinds of jets in HYMORS; consequently, the existence of
FR-dichotomy as a whole is difficult to reconcile with the class of
explanations that posit fundamental differences in the central engine.

}

\maketitle

\section{Introduction}

\citet{fr74} have shown that there are two distinct morphological classes of
double radio galaxies and that there is a relatively sharp transition across
a critical radio luminosity $L_R^{\star}$, corresponding to
$L_{178\,MHz}\simeq5\times10^{24} h^{-2}$ W Hz$^{-1}$ sr$^{-1}$. Most sources
below this luminosity threshold belong to the FR\,I class of radio galaxies.
FR\,I structure consists of diffuse radio lobes that have their brightest
regions within the inner parts of the radio source. The cores are usually
prominent in FR\,I sources and they often show bent jets; in these cases FR\,I
objects can be labelled the Wide Angle Tail, Narrow Angle Tail, and head-tail
sources. More powerful FR\,II-type double sources have their hotspots at the
outer edges of the two radio lobes, and typically only one jet is clearly
detectable. The cores of these sources are weak or hardly observable. There is
strong evidence that FR\,I and FR\,II sources undergo different cosmological
evolution \citep{wall80}; i.e. the density and/or luminosity evolution rate of
FR\,II-type is high, while that of FR\,I appears to be low. It is to be noted
that $L_R^{\star}$ is found to be near a transition in the properties of
nuclear optical emission lines \citep{hl79}.

The origin of FR\,I/FR\,II-dichotomy is much debated in the
astrophysics of extragalactic radio sources. There are three main ways of
interpreting it: 
\begin{itemize}
\item morphological differences are related to the transition of an initially
supersonic but relatively weak jet to a trans\-sonic/subsonic flow
substantially decelerated by thermal plasma within the inner ($\sim$1\,kpc)
region of the host giant elliptical galaxy \citep{k94, blk96, bick95, ka97};
\item there are more fundamental differences between the two classes,
involving the nature of the central black hole and the composition of
jets, i.e. $e^{-} - e^{+}$ plasma for FR\,I sources, while $e^{-} - p$
could be preferred in the case of FR\,II sources
\citep{r96a, r96b, meier97, meier99, cel97};
\item the differences in the jet power/thrust which, together with the
properties of circumgalactic medium, would determine how soon the jets'
collimation is disrupted \citep{gkw88, gk91, gkwh96}.
\end{itemize}

\citet{gkw2000} -- hereafter G-KW -- have introduced a new group of double
radio sources called HYbrid MOrphology Radio Sources (HYMORS), where the two
lobes exhibit clearly different FR morphological types. They argue that HYMORS
could be used to constrain the theoretical models proposed for the
FR-dichotomy, namely, that the existence of HYMORS supports the interpretation
based upon jet interaction with the external medium or -- at least -- that
the models based upon the differences in the nature of central massive black
holes or jet compositions do not seem to be viable, since it is difficult to 
assume a physical process in which the same central engine could produce two 
different types of jets.

The six HYMORS in the sample established by G-KW were selected as a result
of a literature search. Thus, their sample is not homogeneous in many respects.
To circumvent this problem, a programme to find HYMORS candidates in one
comprehensive survey using a single selection procedure was undertaken. This
procedure, the observations, and the data reduction process are described in
Sect.~\ref{sel_obs_red}. The properties of three sources that eventually
turned out to fulfill the criteria of the HYMORS class and two others that are
likely to fall into this category are covered in Sect.~\ref{notes} and
further discussed in Sect.~\ref{disc}.

\begin{table}
\caption {Positions of the centres of subareas searched for HYMORS}
\begin{center}
\begin{tabular}{@{}c c c@{}}
\hline
\hline
\multicolumn{1}{c} { } &
\multicolumn{2}{c} {Position of the centre(J2000)}\\
\cline{2-3}
Subarea & R.A. & Dec.\\
\hline
1 & $10^{\rm h}00^{\rm m}$ & $15\degr00\arcmin$\\
2 & $10^{\rm h}30^{\rm m}$ & $30\degr00\arcmin$\\
3 & $11^{\rm h}00^{\rm m}$ & $45\degr00\arcmin$\\
4 & $12^{\rm h}30^{\rm m}$ & $50\degr00\arcmin$\\
5 & $13^{\rm h}30^{\rm m}$ & $35\degr00\arcmin$\\
\hline
\end{tabular}
\end{center}
\label{t-subareas}
\end{table}

\section{Sample selection, observations, and data reduction}\label{sel_obs_red}

The high sensitivity and the resolution of the VLA {\it Faint Images of the
Radio Sky at Twenty-centimeters} (FIRST) survey
\citep{wbhg97}\footnote{Official website: http://sundog.stsci.edu} offers a
unique possibility of studying the morphologies of a large number of moderately
weak radio galaxies, so it is also a suitable database for pursuing
the search for HYMORS. A high galactic latitude area limited by
RA=$<11^{h}-14^{h}>$ and Dec=$<15\degr-50\degr>$ was inspected with the aim of
finding sources that might belong to the HYMORS class. Our survey did
not completely cover the area indicated above, but instead five subareas
randomly located inside
that region were chosen. Each subarea has a radius of 8\degr20\arcmin
(30000\arcsec). The positions of the centres of these subareas are given in
Table~\ref{t-subareas}.

As a first step, all sources with flux densities
$F_{1.4\,{\rm GHz}}\geq20$\,mJy and angular size $\theta>8\arcsec$ were
selected. Those values were chosen because we preferred to search for
relatively bright extended sources. Maps of more than 1700 sources extracted
from FIRST were inspected visually and the candidates selected. As a result,
a sample consisting of 21~sources that appear to be HYMORS-like in the FIRST
images was established.

This initial candidate sample was observed on 11~Nov. 2003 with the VLA in
B-conf. at 4.9\,GHz, which yields a resolution of $\sim$1\farcs4. Each
programme source was observed for 5.5 min. 3C286 was used as a flux density
calibrator, and sources selected from the {\it List of the VLA calibrators}
nearby to target objects were used as their respective phase calibrators.
The sources were grouped into a few
subsamples according the positions of the targets on the sky, and one phase
calibrator was used for each subsample. The data were reduced using the AIPS
package. After the initial amplitude calibration, a few cycles of
self-calibration were applied. For brighter sources, the amplitude
self-calibration was also carried out. The corrected data were further
processed using IMAGR.

As a result of a careful inspection of 21 VLA 4.9-GHz images,
16~sources were rejected. The reasons for this are described in
Appendix~A. The most common reason is the lack of a well-defined
component that terminates the FR\,II lobe and as such could be
responsible for the edge brightening. In many such cases the lobes are
generally diffuse and weak. Optionally, sources of this kind can be
featured by a (relatively) strong core. If this is the case, the object is
likely to have undergone re-ignition of the activity \citep{mtmb05}. On
the other hand, the lack of a dominating core, together with a ``fuzzy''
shape of the lobes, is a good signature of the cessation of activity. Sources
possessing these features are sometimes termed ``faders'' and, although
relatively rare, have been observed mostly in surveys of ultra-steep spectrum
sources \citep{rot94, db00, coh04}. Six sources out of 16~rejects seem to be
faders, and four others are likely to be restarted. As the images of these
rejected sources might be considered interesting {\em per se}, we include them
as Fig.\,A.1.

The remaining 5 out of 21~sources observed at 4.9\,GHz using the VLA
appeared to be HYMORS according to the same criteria
as those adopted by G-KW. The follow-up observations of these objects were
carried out with the VLA in A-conf. at 1.4\,GHz on 20 and 21~Sep. 2004.
Again, 3C286 was used as the flux density calibrator, and programme sources
were grouped so that a single phase calibrator could be used for each group.
Each target source was observed for $\sim$30\,min. divided into six
$\sim$5-min. scans.

\begin{figure*}
\centering
\includegraphics[scale=0.44]{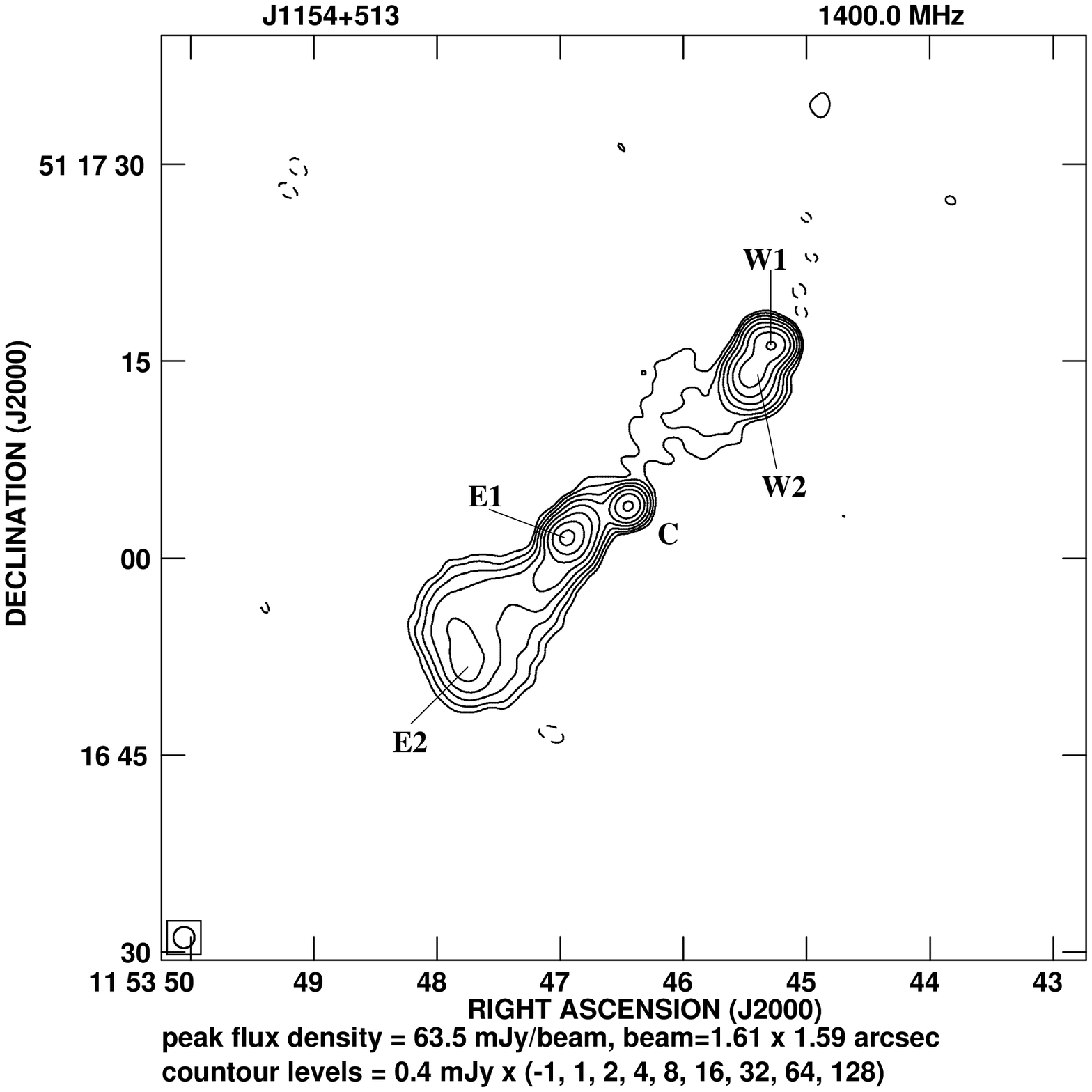}
\includegraphics[scale=0.44]{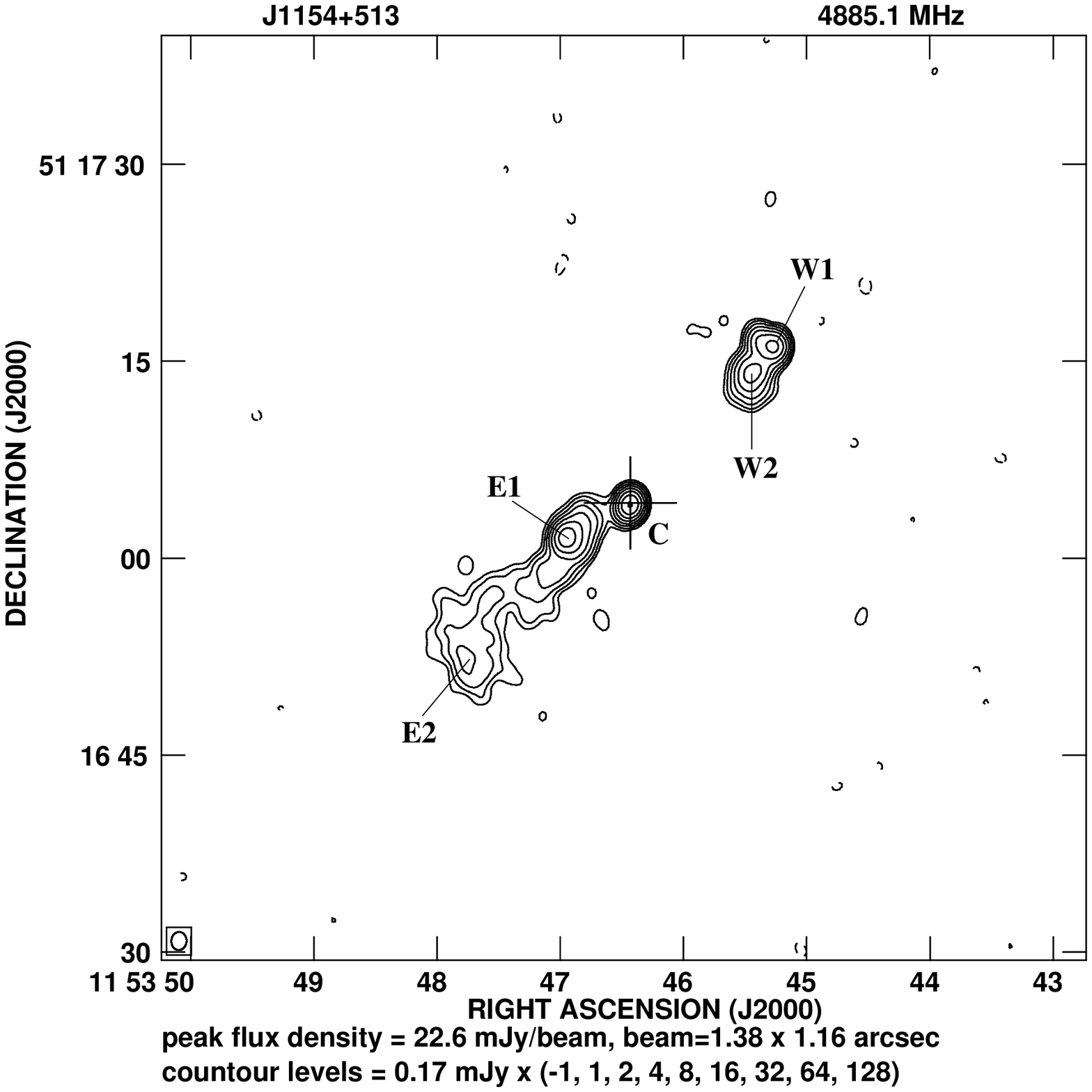}
\includegraphics[scale=0.44]{3996f1c.ps}
\includegraphics[scale=0.44]{3996f1d.ps}
\caption[]{VLA maps of {\bf J1154+513}. \emph{Upper left}: VLA in A-conf.
at 1.4\,GHz. \emph{Upper right}: VLA in B-conf. at 4.9\,GHz. The position
of the optical object extracted from SDSS is marked with a cross. \emph{Lower
left}: FIRST map. \emph{Lower right}: Spectral index map obtained from the
two-frequency VLA images shown in the upper panels.}
\label{f-J1154+513}
\end{figure*}

\begin{figure*}
\centering
\includegraphics[scale=0.44]{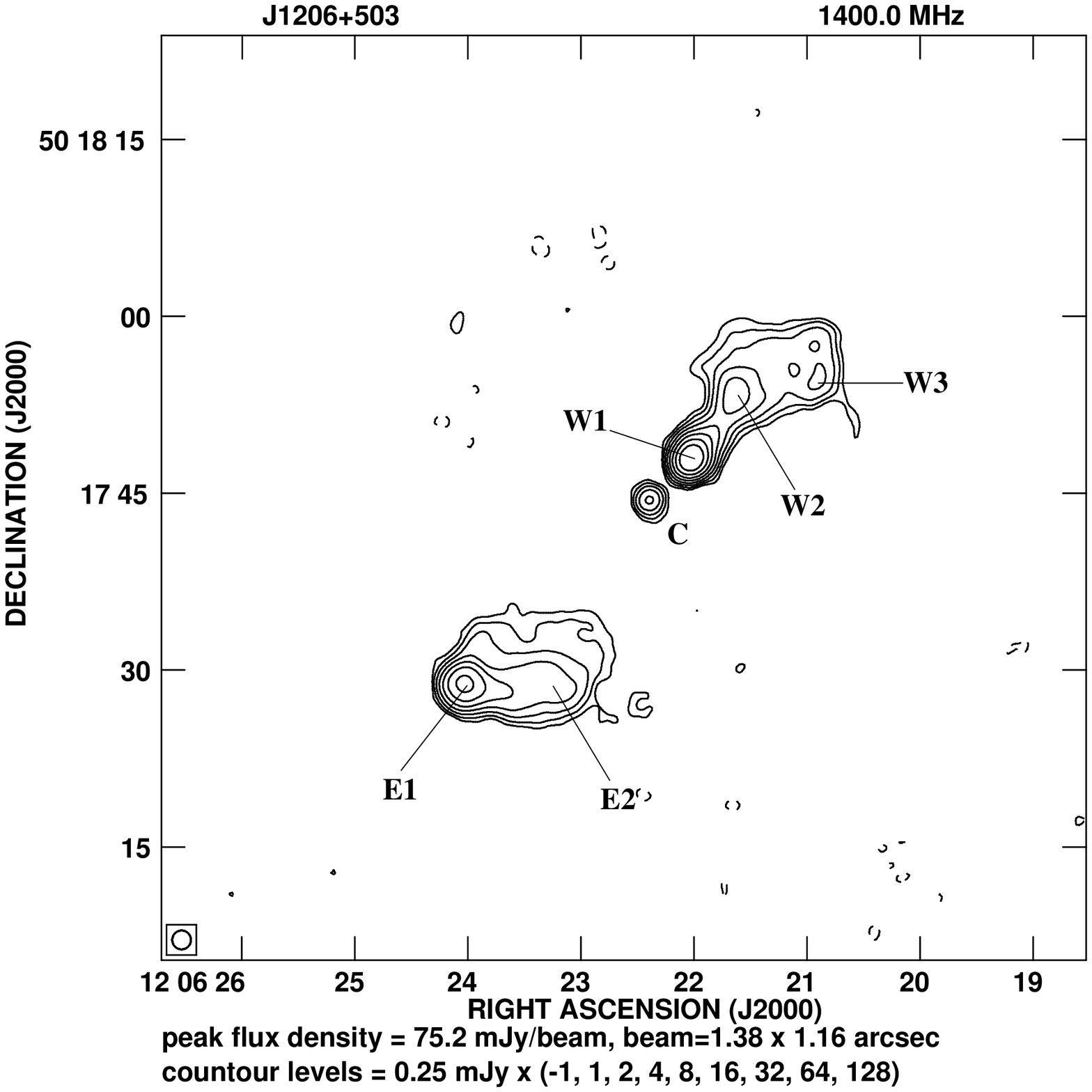}
\includegraphics[scale=0.44]{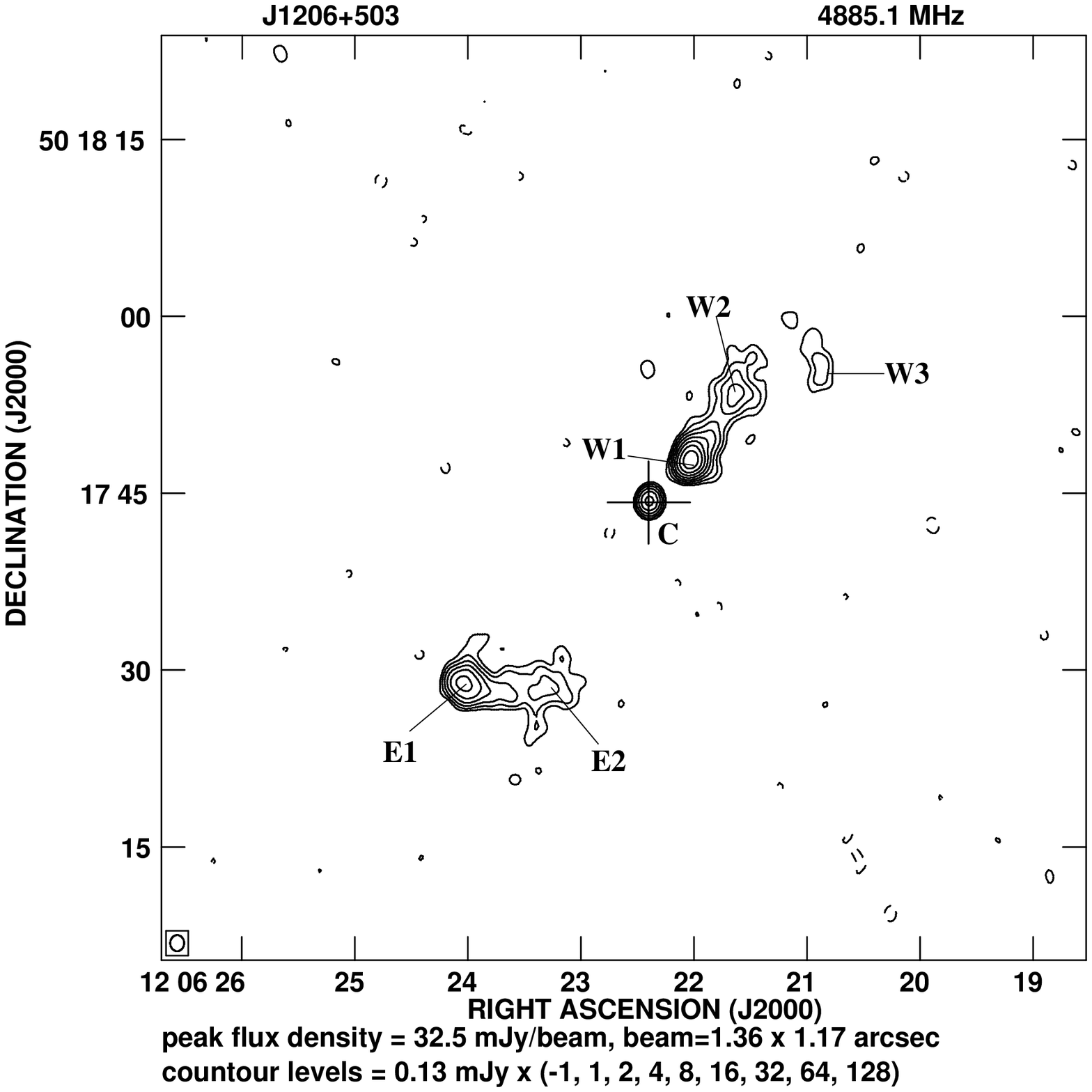}
\includegraphics[scale=0.44]{3996f2c.ps}
\includegraphics[scale=0.42]{3996f2d.ps}
\caption[]{VLA maps of {\bf J1206+503}. \emph{Upper left}: VLA in A-conf.
at 1.4\,GHz. \emph{Upper right}: VLA in B-conf. at 4.9\,GHz. The position
of the optical object extracted from SDSS is marked with a cross. \emph{Lower
left}: FIRST map. \emph{Lower right}: Spectral index map obtained from the
two-frequency VLA images shown in the upper panels.}
\label{f-J1206+503}
\end{figure*}

\begin{figure*}
\centering
\includegraphics[scale=0.44]{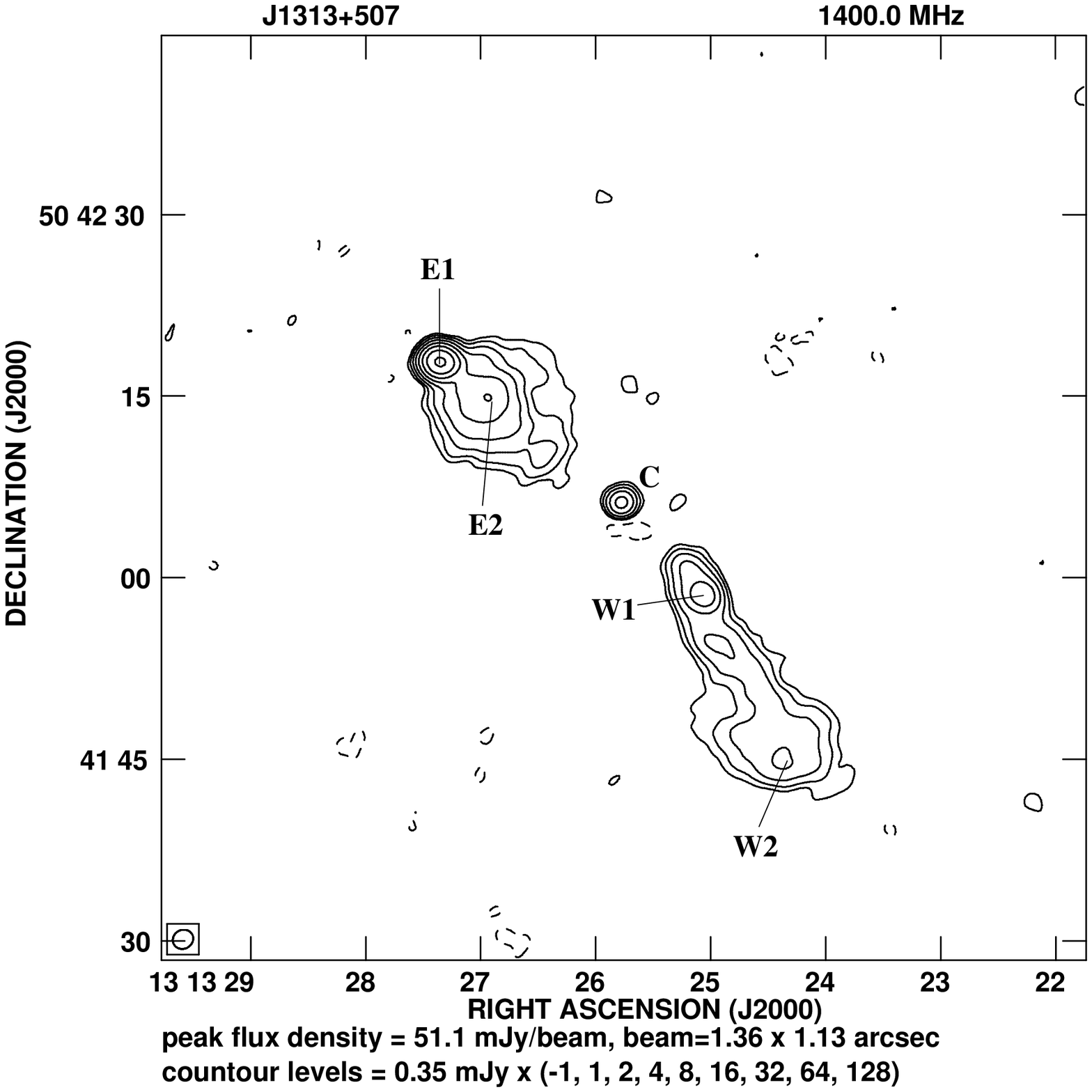}
\includegraphics[scale=0.44]{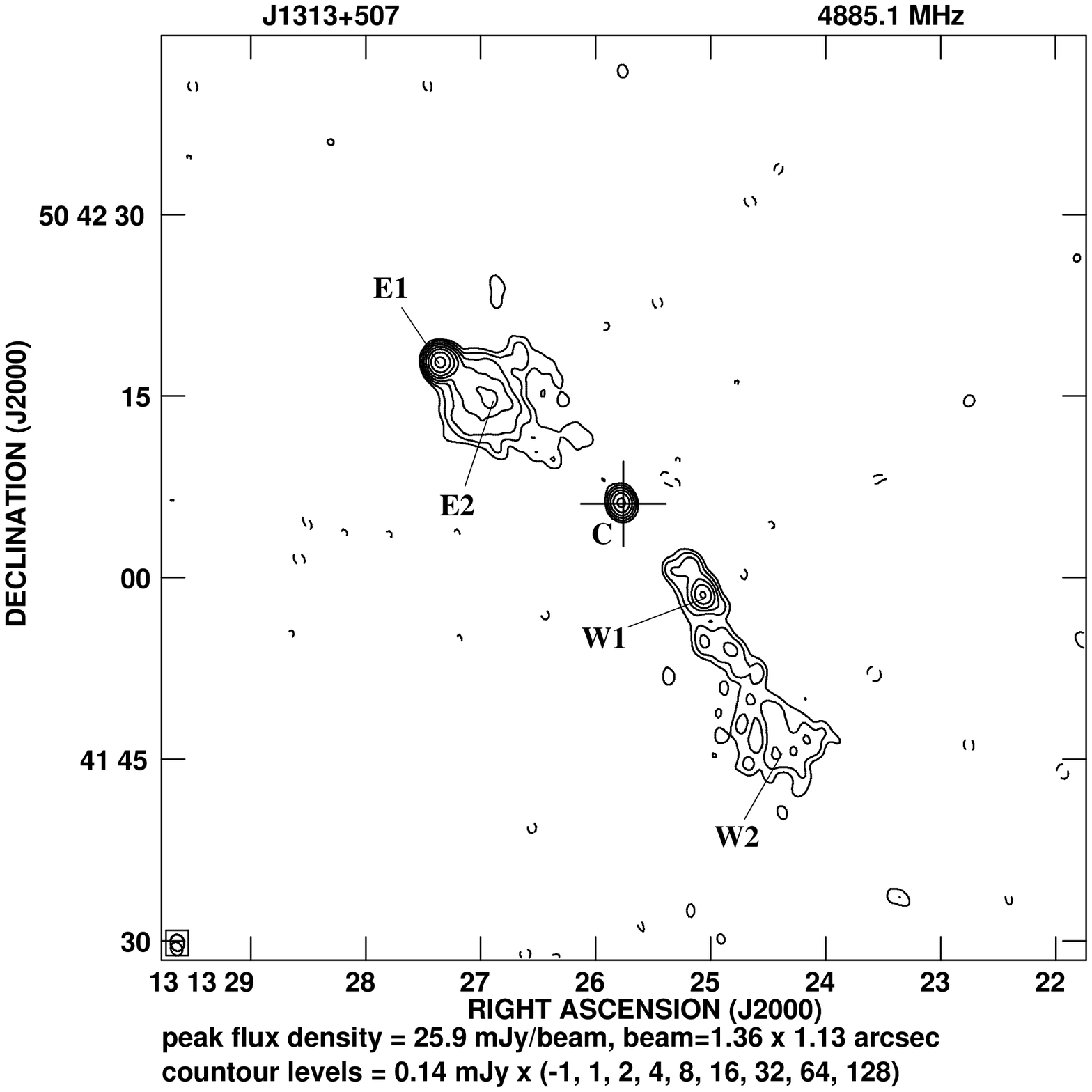}
\includegraphics[scale=0.44]{3996f3c.ps}
\includegraphics[scale=0.44]{3996f3d.ps}
\caption[]{VLA maps of {\bf J1313+507}. \emph{Upper left}: VLA in A-conf.
at 1.4\,GHz. \emph{Upper right}: VLA in B-conf. at 4.9\,GHz. The position
of the optical object extracted from SDSS is marked with a cross. \emph{Lower
left}: FIRST map. \emph{Lower right}: Spectral index map obtained from the
two-frequency VLA images shown in the upper panels.}
\label{f-J1313+507}
\end{figure*}

\begin{figure*}
\centering
\includegraphics[scale=0.44]{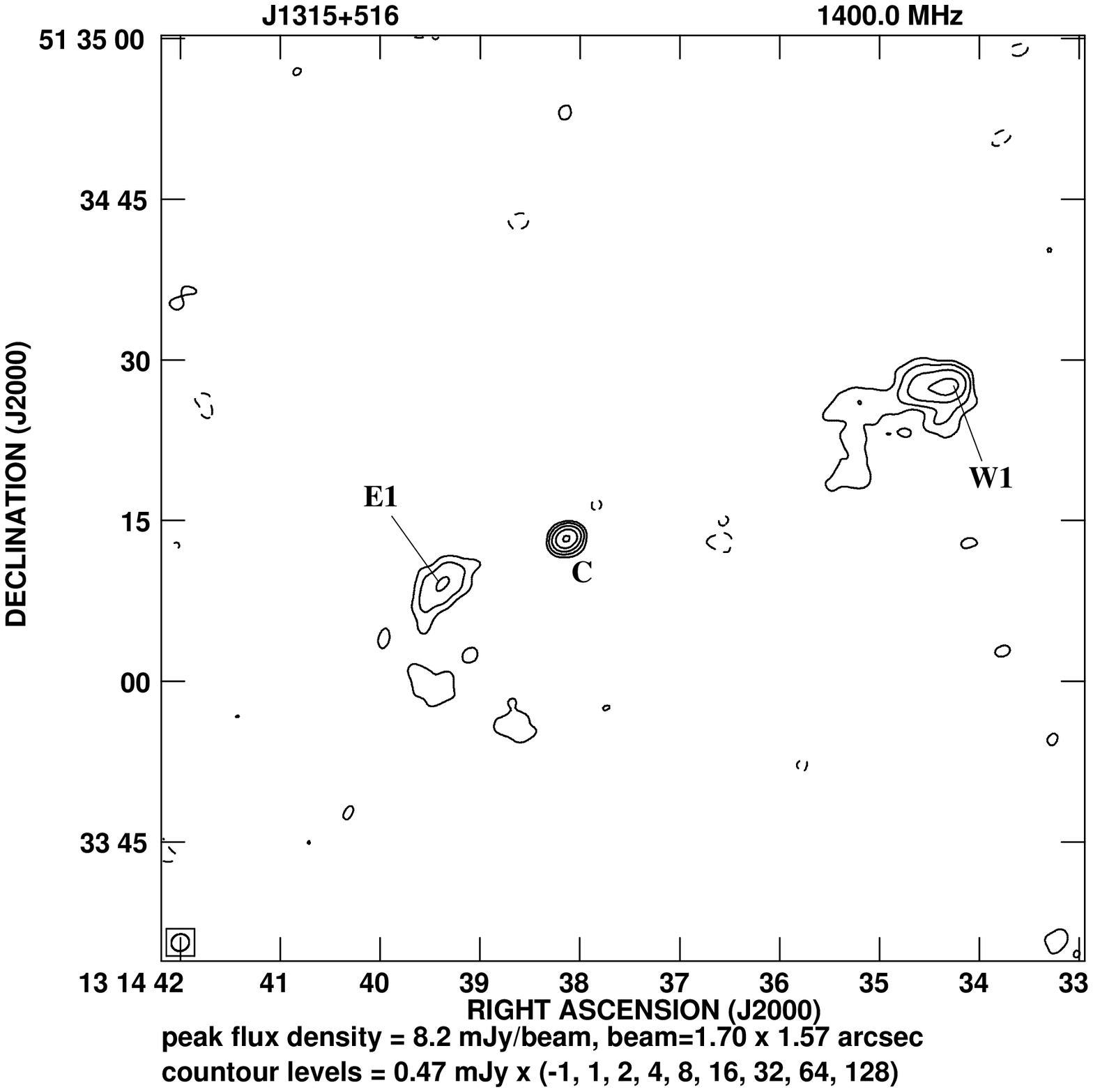}
\includegraphics[scale=0.44]{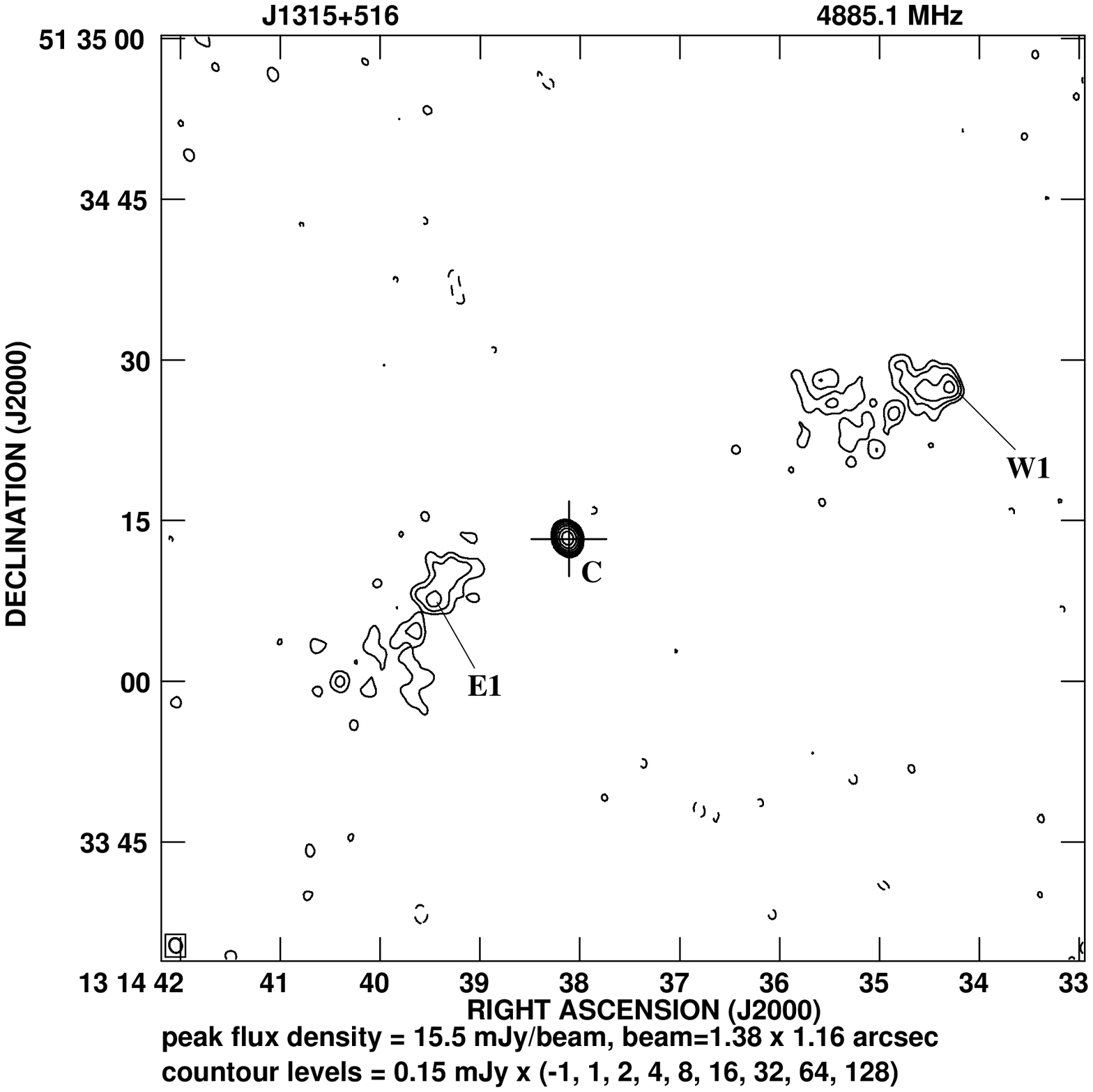}
\includegraphics[scale=0.44]{3996f4c.ps}
\caption[]{VLA maps of {\bf J1315+516}. \emph{Upper left}: VLA in A-conf.
at 1.4\,GHz. \emph{Upper right}: VLA in B-conf. at 4.9\,GHz. The position
of the optical object extracted from SDSS is marked with a cross. \emph{Lower}:
FIRST map.}
\label{f-J1315+516}
\end{figure*}

\begin{figure*}
\includegraphics[scale=0.44]{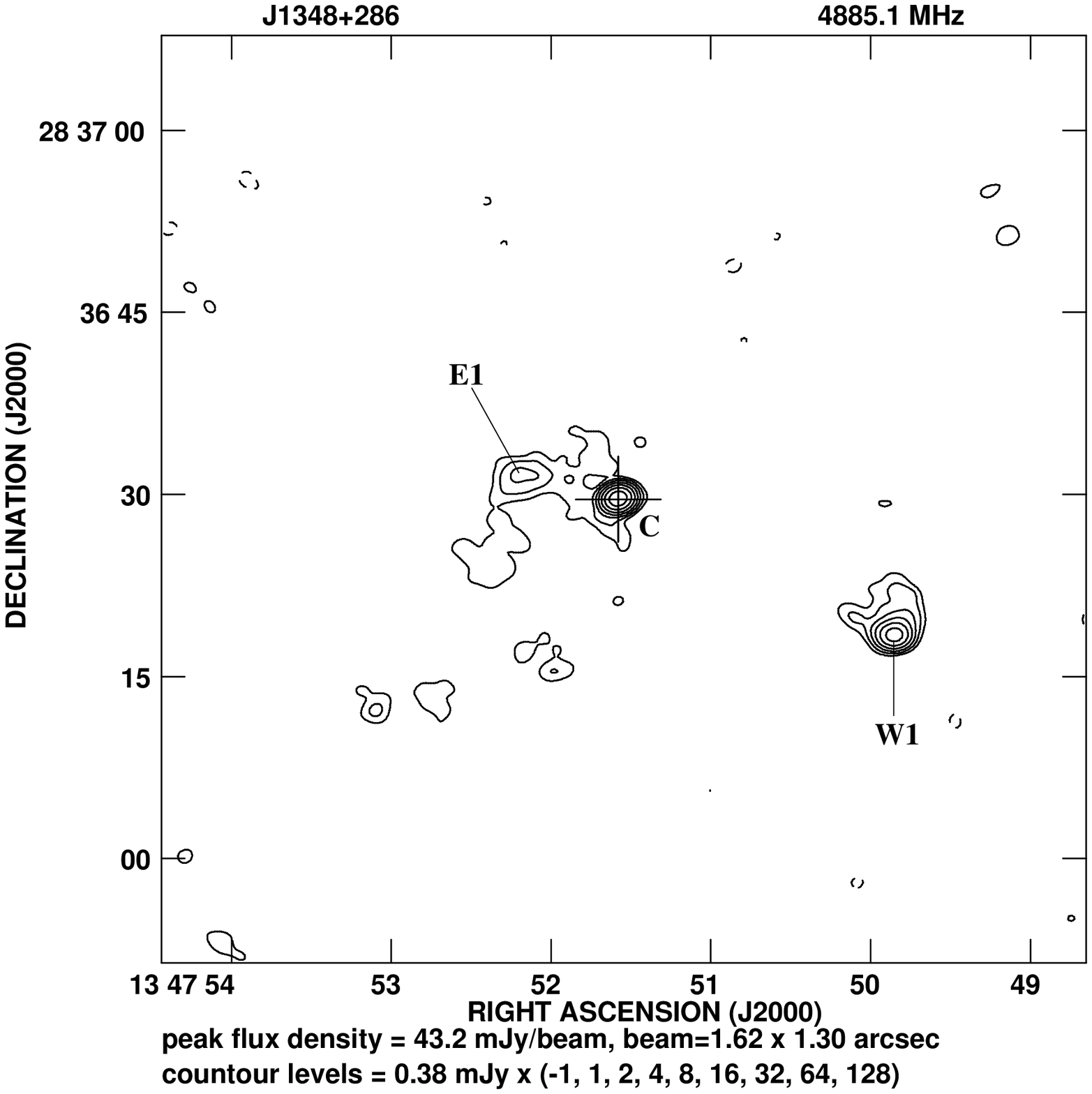}
\includegraphics[scale=0.44]{3996f5b.eps}
\caption[]{VLA maps of {\bf J1348+286}. \emph{Left}: VLA in B-conf.
at 4.9\,GHz. The position of the optical object extracted from SDSS is marked
with a cross. \emph{Right}: FIRST map.}
\label{f-J1348+286}
\end{figure*}

The basic parameters of the new HYMORS are given in Table~\ref{t-basicdata},
and the final images shown in Figs.~\ref{f-J1154+513} to~\ref{f-J1348+286}.
For comparison, their respective FIRST images are also included. The positions
listed in Table~\ref{t-basicdata} are those of the core components as seen in
the 4.9-GHz VLA maps fitted using AIPS task JMFIT. JMFIT was also used to
measure the flux densities of the other main components, and these measurements
are shown in Table~\ref{t-fluxes}. For three sources (J1154+513, J1206+503 and
J1313+507), spectral index maps were obtained from the two-frequency VLA images
convolved with a common circular Gaussian beam 
($1\farcs75\times 1\farcs75$) and are shown in Figs.~\ref{f-J1154+513}
to~\ref{f-J1313+507}. As all five sources investigated here are included in the
Release 4 of the Sloan Digital Sky Survey (SDSS/DR4)\footnote{The up-to-date
version at the time of writing.}, their SDSS objID's are listed in
Table~\ref{t-basicdata} (column 2).

\begin{table*}
\caption {The newly discovered HYMORS. Positions are those of the core
components.}
\begin{tabular}{@{}l l c c c c c r c@{}}
\hline
\hline
\multicolumn{4}{c} {} &
\multicolumn{5}{c} {Flux density [mJy]} \\
\cline {5-9}
\multicolumn{1}{c} {Source} &
\multicolumn{1}{c} {SDSS} &
\multicolumn{1}{c} {R.A.} &
\multicolumn{1}{c} {Dec.} &
\multicolumn{3}{c} {1.4\,GHz} &
\multicolumn{2}{|c} {4.9\,GHz} \\
\cline {3-9}
\multicolumn{1}{c} {name} &
\multicolumn{1}{c} {objID} &
\multicolumn{2}{c} {(J2000)} &
\multicolumn{1}{c} {FIRST} &
\multicolumn{1}{c} {NVSS} &
\multicolumn{1}{c} {this paper} &
\multicolumn{1}{|c} {GB6} &
\multicolumn{1}{c} {this paper} \\
\hline
J1154+513 & 587732134310838583 & ~11~ 53~ 46.43~ & ~+51~ 17~ 04.1~ & ~495~ & ~483~ & ~490~ & ~137~ & ~131~ \\
J1206+503 & 587732483822518553 & ~12~ 06~ 22.39~ & ~+50~ 17~ 44.3~ & ~241~ & ~170~ & ~265~ & ~~75~ & ~~88~ \\
J1313+507 & 588018054571360837 & ~13~ 13~ 25.78~ & ~+50~ 42~ 06.2~ & ~277~ & ~252~ & ~230~ & ~~84~ & ~~86~ \\
J1315+516 & 588018055645102348 & ~13~ 14~ 38.12~ & ~+51~ 34~ 13.4~ & ~~93~ & ~144~ & ~~~48~ & ~~~51~ & ~~~41~ \\
J1348+286 & 587739721369255946 & ~13~ 47~ 51.58~ & ~+28~ 36~ 29.6~ & ~241~ &       &       & ~117~ & ~105~ \\

\hline
\end{tabular}

\vspace {0.5cm}

\label{t-basicdata}
\end{table*}

\section{Notes on individual sources}\label{notes}

{\bf\object{J1154+513}} (Fig.~\ref{f-J1154+513}). In the FIRST image, this
object (also known as 4C\,+51.28) appears as a triple. There is a hint that the
south-eastern component, together with the central feature could form an
unresolved FR\,I jet. This conjecture was fully confirmed by our VLA
observations at 4.9\,GHz and 1.4\,GHz. According to the images resulting from
them, it is evident that the south-eastern part of the source has a classical
FR\,I structure with flat spectrum ($\alpha=-0.25$)\footnote{Throughout this
paper $\alpha$ is defined as: $S\propto\nu^{\alpha}$.} bright core ``C'' and a
prominent hotspot ``E1'' placed inside the jet. On the other hand, the
north-western part is a typical FR\,II lobe (``W2'') with a hotspot (``W1'').
The ``E2'' component has a steeper spectrum ($\alpha=-1.45$) than E1
($\alpha=-1.05$). Steepening of the spectrum with the increasing distance from
the core is a well-known feature of FR\,I radio galaxies
\citep[see e.g.][]{ksr97}. The north-western lobe has a structure that is
typical for FR\,II-type lobes with
flattening of spectrum toward the hotspot \citep[see e.g.][]{lms89}. Therefore,
both the morphology, as seen in the total intensity maps, and the
spectral index map fully confirm a hypothesis that there are two
different types of jets present in one object. It is also to be noted that
there is a connection (bridge) between the northern FR\,II-type lobe and the
core ``C'' clearly visible at 1.4\,GHz. Given that it is not present in the
4.9-GHz image, the bridge must have a very steep spectrum. This remains in full
agreement with \citet{lms89} leaving no room for doubt that the two parts of
the source as seen in the 4.9-GHz VLA image are not a coincidence. Our
measurements of the flux densities are in a very good agreement with GB6
(4.85\,GHz) and NVSS (1.4\,GHz) catalogues. This source is very similar to
\object{1004+130} and \object{4C\,$-$03.64}, both presented in G-KW, and as
such is a perfect example of a HYMORS.

The J1154+513 radio source has an optical counterpart in SDSS/DR4, a
galaxy with $m_R=21.46$. There is a hint of irregularity in the shape of the
host galaxy. The redshift of this object is unknown at present.

{\bf\object{J1206+503}} (Fig.~\ref{f-J1206+503}). In the FIRST image this source
has two main components. The south-eastern one is an FR\,II-type lobe, whereas
the north-western part of source has a barely resolved core-jet structure. The
images resulting from our VLA follow-up observations confirm this. The
north-western part consists of three conspicuous features, a flat spectrum core
(``C'') and the jet with two brighter knots (``W1'' and ``W2'') and a diffuse
``plume'' (``W3''), which together make a clear case for a FR\,I-type jet,
whereas the eastern lobe (``E2'') with a hotspot (``E1'') is an FR\,II-type
lobe specimen. Thus, the morphology of J1206+503 is fully consistent with a
HYMORS class definition. As in the case of J1154+513, the spectral index map of
J1206+503 confirms that there are two different types of radio jets.

This source has an $m_R=20.86$ optical counterpart in SDSS/DR4. The host galaxy
seems to be irregular and has two close companion galaxies:
SDSS\,J120622.11+501740.9 ($m_R=22.66$) and SDSS\,J120621.94+501743.2
($m_R=21.21$) located 4\farcs3 and 4\farcs6 off the position of the optical
object identified with the core of the HYMORS, respectively. The redshifts of
all three objects are not known at present.

{\bf\object{J1313+507}} (Fig.~\ref{f-J1313+507}). This source consists of four
main components in the FIRST image. A well-developed north-eastern FR\,II
structure, a putative core with an optical counterpart -- $m_R=20.90$ in
SDSS/DR4 -- and a south-western structure that could be an unresolved FR\,I jet.
Our new VLA images confirm that the latter is actually FR\,I-type. With the
``E1'' component being the hotspot of an FR\,II-type lobe and ``W2'' FR\,I-type
plume, J1313+507 is -- again -- a very good example of a HYMORS, quite similar
to 4C\,$-$03.64 (G-KW). The redshift of this object is unknown at present.

{\bf\object{J1315+516}} (Fig.~\ref{f-J1315+516}). The FIRST image strongly
suggests that this source has a HYMORS morphology. In the 4.9-GHz image, the
eastern (apparently FR\,I) jet is not reproduced well, which means it has a
diffuse structure without any well-localised features except the region denoted
as ``E1''. The western part
consists of a hotspot (``W1'') inside a well-defined FR\,II-like lobe. Given
that a substantial amount of flux is missing when observing the source at
1.4\,GHz in B-conf. (FIRST) and A-conf. (our observations), a significant
diffuse component must exist that is not reproduced in either our 1.4-GHz
image or the FIRST map. Also, the weak radio structures in this object do not
allow a good map of the spectral index to be made. The fact that the source
is strongly core-dominated, whereas the lobes are very diffuse, is an
indication that it could be a restarted source -- see Sect.~\ref{disc} for
further discussion.

The core of J1315+516 has an optical counterpart of $m_R=19.77$, and there are
two close companion galaxies: SDSS\,J131437.79+513410.6 ($m_R=22.76$) and
SDSS\,J131437.10+513408.1 ($m_R=20.70$) located 4\farcs1 and 10\farcs8
off the position of the optical object identified with the core of the HYMORS.
The redshifts of all these three objects are not known at present.

{\bf\object{J1348+286}} (Fig.~\ref{f-J1348+286}). The structure of this object
is somewhat similar to that of J1315+516: a hotspot of the FR\,II-like lobe
(``W1''), and a fuzzy, possible FR\,I jet (``E1'') can be recognised. Due to
the bad quality of the observational data, it was not possible to make a good
VLA image of this source at 1.4\,GHz, so the FR\,I nature of the ``E1''
structure is not certain, and this is the least convincing HYMORS among the
five shown here.

J1348+286 is associated with an X-ray object RX\,J1347.7+2836. According to
SDSS/DR4 this object is a QSO ($m_R=17.27$) at the redshift of $z=0.7407$
\citep{mun03}. (A $m_R=14.5$ field star is within 2\farcs7 the QSO.) This is
the farthest HYMORS known up to date. At this redshift and given that the total
flux of the source extracted from FIRST is 241\,mJy, the logarithm of the
monochromatic luminosity at 1.4\,GHz amounts to 26.5, assuming
$H_0$=75\,km\,s$^{-1}$\,Mpc$^{-1}$ and $q_0=0.5$ in accordance with the data
in Table~1 in G-KW. It looks, therefore, that this distant HYMORS is an order
of magnitude more powerful than the nearby ones examined by G-KW ($\log L_R =
25.4$ for three of them), although it must be borne in mind that for 1004+130
$\log L_R = 26.3$, which is quite close to the respective value for J1348+286.

\begin{table}
\caption {The flux densities and spectral index of main components}
\begin{tabular}{l l r r c}
\hline
\hline
\multicolumn{1}{l} {Source} &
\multicolumn{1}{l} {Com-} &
\multicolumn{2}{c} {Flux density [mJy]} &
\multicolumn{1}{c} {Spectral} \\
\cline{3-4}
& 
\multicolumn{1}{l} {ponent} &
\multicolumn{1}{c} {1.4\,GHz} &
\multicolumn{1}{c} {4.9\,GHz} &
\multicolumn{1}{c} {index} \\
\hline
J1154+513 &  C  &  33.4 & 24.7 & $-$0.25 \\
          &  E1 & 107.5 & 29.3 & $-$1.05 \\
          &  E2 &  54.0 &  9.1 & $-$1.45 \\
          &  W1 & 100.9 & 22.0 & $-$1.24 \\
          &  W2 &  99.4 & 18.6 & $-$1.36 \\
\hline
J1206+503 &  C &   4.5 &   5.8 & ~~0.21 \\
          &  E1 &  52.1 &  15.5 & $-$0.99 \\
          &  E2 &  30.9 &   4.6 & $-$1.54 \\
	        &  W1 & 105.9 &  45.9 & $-$0.68 \\
          &  W2 &  26.5 &   5.2 & $-$1.32 \\
	        &  W3 &   9.4 &   1.4 & $-$1.56 \\
\hline
J1313+507 &  C  &   7.0 &   6.1 & $-$0.11 \\
          &  E1 &  69.8 &  30.8 & $-$0.66 \\
          &  E2 &  58.0 &  15.3 & $-$1.08 \\
          &  W1 &  17.9 &   6.7 & $-$0.80 \\
\hline
J1315+516 &  C  & 16.0 & 13.1 &  ~~0.09 \\
          &  E1 &  3.1 & 21.7 & $-$0.55 \\
          &  W1 &  3.6 & 19.9 & $-$0.58 \\
\hline
J1348+286 &  C  &      & 44.7 &       \\
          &  E1 &      &  8.7 &       \\
          &  W1 &      & 24.7 &       \\

\hline
\end{tabular}
\label{t-fluxes}
\end{table}

\section{Discussion}\label{disc}

HYMORS could serve as a discriminator between a range of theories used to
explain the origin of the FR-dichotomy. As pointed out by G-KW, the existence
of HYMORS does not favour the models based upon fundamental differences between
the central engines, such as black hole spin or jet composition. It seems that
in the case of HYMORS some type of jet-medium interaction on the scale of
kiloparsecs may play a crucial role and has a significant impact on the
FR-dichotomy. Thus, if it is assumed that the properties of the intracluster 
or intergalactic
medium are an important factor in the evolution of radio sources, then the
phenomenon of HYMORS might be explained. It is likely that, during the
evolution of a cluster of galaxies, there are many interactions between the
members of the cluster. If the interactions are very frequent and strong, it
may be expected that the properties of the intracluster medium (density,
temperature, pressure) in the cluster could vary in time and space. Assuming
that there is a difference of the properties in the medium between the opposite
sides of a host galaxy, a HYMORS might emerge even if the jets are identical
when launched from the central engine. A recent study by \cite{cr04} also
provides support for the hypothesis that there is no fundamental difference
between FR\,I and FR\,II engines.

It is to be noted that three sources out of the five investigated in detail --
J1154+513, J1206+503 and J1313+507 -- are not core-dominated (see
Table~\ref{t-fluxes}), and their structures are imaged very well using the
VLA with 1\farcs4 resolution, whereas the remaining two -- J1315+516 and
J1348+286 -- are clearly core-dominated, but their jets, particularly those
at the ``FR\,I side'', are fuzzy and not reproduced well in our maps, as
substantial amounts of the flux are missing. We suggest that these two objects,
apart from being likely examples of HYMORS, can also be labeled as
``core-dominated triples''. Radio sources of this sort extracted from FIRST in
a systematic manner have been investigated by \citet{mtmb05}. They claim that
such objects are restarted and that the dominance of their cores is a direct
consequence of being re-oriented in the course of a merger event. As a result
of this, they are now more beamed towards the observers whereas the lobes are
not fuelled any longer, so they have entered the so-called ``coasting'' phase
of their evolution. However, if these two are not restarted sources, then
the core prominence would imply that both lobes are greatly foreshortened by
projection, in which case the apparent FR\,I structure could be an FR\,II-type
lobe seen nearly head-on.

\section{Summary}

The main objective of this work was to expand the number of known HYMORS by
means of a systematic study of images resulting from a high sensitivity radio
sky survey. A sample of more than 1700~sources from the FIRST catalogue was
examined and 21~candidates selected. After re-observation with the VLA in
B-conf. at 4.9\,GHz, three sources turned out to be certain HYMORS and two
others are very likely to fall into this rare category. Our ``success rate''
therefore is somewhat lower than that of G-KW, as they found 6~HYMORS among
the somewhat more than 1000~objects they examined from a search of the
literature. It must be stressed, however, that in the case of the FIRST
catalogue, the resolution of the images is often worse than those in the images
selected by G-KW. As a result we might well have rejected some true HYMORS
during the initial selection process. Follow-up VLA observations in A-conf. at
1.4\,GHz provided a confirmation that those three objects labelled as HYMORS
based upon the inspection of 4.9-GHz images indeed fulfill the criteria of
this class. They also made the preparation of spectral index maps possible.
The spectral index gradients additionally support the identifications of the
FR\,I and FR\,II sides. As a result, the conjecture that HYMORS indeed contain
two different types of radio jets has gained strong support.

\begin{acknowledgements}

\item The VLA is operated by the U.S. National Radio Astronomy Observatory,
which is operated by Associated Universities, Inc., under cooperative
agreement with the National Science Foundation.

\item This research has made use of the NASA/IPAC Extragalactic Database
(NED), which is operated by the Jet Propulsion Laboratory, California
Institute of Technology, under contract with the National Aeronautics and
Space Administration.

\item Use has been made of the fourth release of the Sloan Digital Sky
Survey (SDSS) Archive. Funding for the creation and distribution of the
SDSS Archive has been provided by the Alfred P. Sloan Foundation, the
Participating Institutions, the National Aeronautics and Space
Administration, the National Science Foundation, the U.S. Department of
Energy, the Japanese Monbukagakusho, and the Max Planck Society. The SDSS
Web site is http://www.sdss.org/. The SDSS is managed by the Astrophysical
Research Consortium (ARC) for the Participating Institutions. The
Participating Insti\-tutions are The University of Chicago, Fermilab, the
Insti\-tute for Advanced Study, the Japan Participation Group, The Johns
Hopkins University, Los Alamos National Labora\-tory, the
Max-Planck-Institute for Astronomy (MPIA), the Max-Planck-Institute for
Astrophysics (MPA), New Mexico State University, University of Pittsburgh,
Princeton University, the United States Naval Observatory, and the
University of Washington.

\item We are very grateful to Prof. Paul Wiita and Prof. Gopal-Krishna for
carefully reading the manuscript and making a number of suggestions
and corrections that led to a significant improvement in this paper.

\end{acknowledgements}

\Online

\section*{Appendix A}

Here we present notes on 16 sources selected from FIRST as possible candidates
for HYMORS, observed at 4.9\,GHz with the VLA and rejected based on the results
of these observations. The 4.9-GHz VLA images of these sources are shown in
Fig.\,A.1 along with their respective cutouts from FIRST. The
positions of the optical counterparts, if available in SDSS/DR4, are marked
with crosses in the 4.9-GHz VLA images. They are also listed in
Table\,A.1.

{\bf\object{J1011+328}}.
No compact feature that might be labelled as a hotspot seen in the VLA 4.9-GHz
image, so there is no clear case of the FR\,II part. This object is a member
of a cluster. The SDSS object 587739157655453889 is a possible core of this
radio source, so that it is a Wide-Angle Tail (WAT) source. Another SDSS object
(587739157655453835) is also within the boundary of the radio image presented,
and it is marked with $\times$. J1011+328 is possibly a fader.

{\bf\object{J1020+141}}.
Although there is a hint in the VLA 4.9-GHz image that this could be a HYMORS,
we label it as an FRI radio galaxy with a distorted structure. This galaxy at a
redshift $z=0.146$ is a member of a cluster and has three close companions.

{\bf\object{J1021+444}}.
No FR\,I part is seen in the VLA 4.9-GHz image.

{\bf\object{J1025+277}}.
No compact feature that might be labelled as a hotspot is seen in the VLA
4.9-GHz image so there is no clear case of the FR\,II part.
There is a lack of visible jets and hotspots, and the radio core is weak
($S_{core}/S_{tot}=$0.052). We suggest this source could be the same
type as a fading galaxy 1855+37 shown by \citet{gir05}. 

{\bf\object{J1029+373}}.
No compact feature that might be labelled as hotspot is seen in the VLA
4.9-GHz image so there is no clear case of the FR\,II part.
As this source is core-dominated, it is likely to be restarted.

{\bf\object{J1030+299}}.
The radio structure shows no indication of hotspots in the faint lobes
or the radio core. This object resembles 1542+323, the fading source presented
by \citet{kms05}.

{\bf\object{J1045+523}}.
This source is a possible fader.

{\bf\object{J1048+153}}.
We failed to make a good 4.9-GHz image. Nevertheless, based on the FIRST image
alone, this source looks like a very good HYMORS candidate.

{\bf\object{J1125+374}}.
As this source is core dominated, it is likely to be restarted.

{\bf\object{J1203+538}}.
The core is weak ($S_{core}/S_{tot}=$0.067), and there is a lack of visible
jets and hotspots. We suggest this source could be the same type as 1855+37,
the fading galaxy shown by \citet{gir05}.

{\bf\object{J1207+483}}.
No compact feature that might be labelled as a hotspot is seen in the VLA
4.9-GHz image, so although this source has a perfect northern FR\,I structure,
there is no clear case of the FR\,II part.

{\bf\object{J1210+466}}.
As this source is core-dominated, it is likely to be restarted.

{\bf\object{J1303+318}}.
No compact feature that might be labelled as a hotspot is seen in the VLA
4.9-GHz image, so there is no clear case of the FR\,II part.
It is a member of the Abell\,1667 cluster.
As this source is core-dominated, it is likely to be restarted.

{\bf\object{J1324+376}}.
In the 4.9-GHz VLA image, the radio structure of this
source consists of a core and diffuse lobes without hotspots, so that it is
impossible to classify this source either as FR\,I or FR\,II. It is a member
of the ZwCl\,1322.0+3750 cluster. J1324+376 is a possible fader.

{\bf\object{J1339+394}}.
FRII radio galaxy with a bright core.

{\bf\object{J1351+309}}.
FRI radio galaxy belonging to the ZwCl\,1348.7+3109 cluster.

\begin{table}
{\small{\bf Table\,A.1.}
Rejected HYMORS candidates with optical counterparts in SDSS/DR4.
Positions are those of the optical object.}
\begin{tabular}{@{}l l c c@{}}
\hline
\hline
\multicolumn{1}{c} {Source} &
\multicolumn{1}{c} {SDSS} &
\multicolumn{1}{c} {R.A.} &
\multicolumn{1}{c} {Dec.} \\
\cline {3-4}
\multicolumn{1}{c} {name} &
\multicolumn{1}{c} {objID} &
\multicolumn{2}{c} {(J2000)} \\
\hline

J1011+328 & 587739157655453889 & 10 11 06.12 & 32 48 17.2 \\
J1020+141 & 587735349636169918 & 10 19 32.33 & 14 03 01.8 \\
J1021+444 & 588017112629313672 & 10 21 16.44 & 44 25 48.1 \\
J1029+373 & 587738948284842142 & 10 28 30.52 & 37 15 10.7 \\
J1045+523 & 587733080269914893 & 10 44 38.63 & 52 15 37.4 \\
J1125+374 & 587739098595328059 & 11 24 38.16 & 37 22 40.3 \\
J1203+538 & 587733079201349782 & 12 03 02.05 & 53 49 18.7 \\
J1207+483 & 588297865268559953 & 12 06 56.81 & 48 15 12.5 \\
J1210+466 & 588297863121273573 & 12 09 47.09 & 46 37 05.1 \\
J1303+318 & 587739505547673753 & 13 03 17.11 & 31 50 02.6 \\
J1324+376 & 587739099142553704 & 13 24 12.40 & 37 33 33.8 \\
J1339+394 & 587738575145861682 & 13 39 08.54 & 39 26 24.1 \\
J1351+309 & 587739609709674629 & 13 51 03.40 & 30 54 04.2 \\
\hline
\end{tabular}
\label{t-sdssids}.
\end{table}

\newpage
\begin{figure*}
\centering
\includegraphics[scale=0.4]{3996fA1a.ps}
\includegraphics[scale=0.4]{3996fA1b.ps}
\includegraphics[scale=0.4]{3996fA1c.ps}
\includegraphics[scale=0.4]{3996fA1d.ps}
\includegraphics[scale=0.4]{3996fA1e.ps}
\includegraphics[scale=0.4]{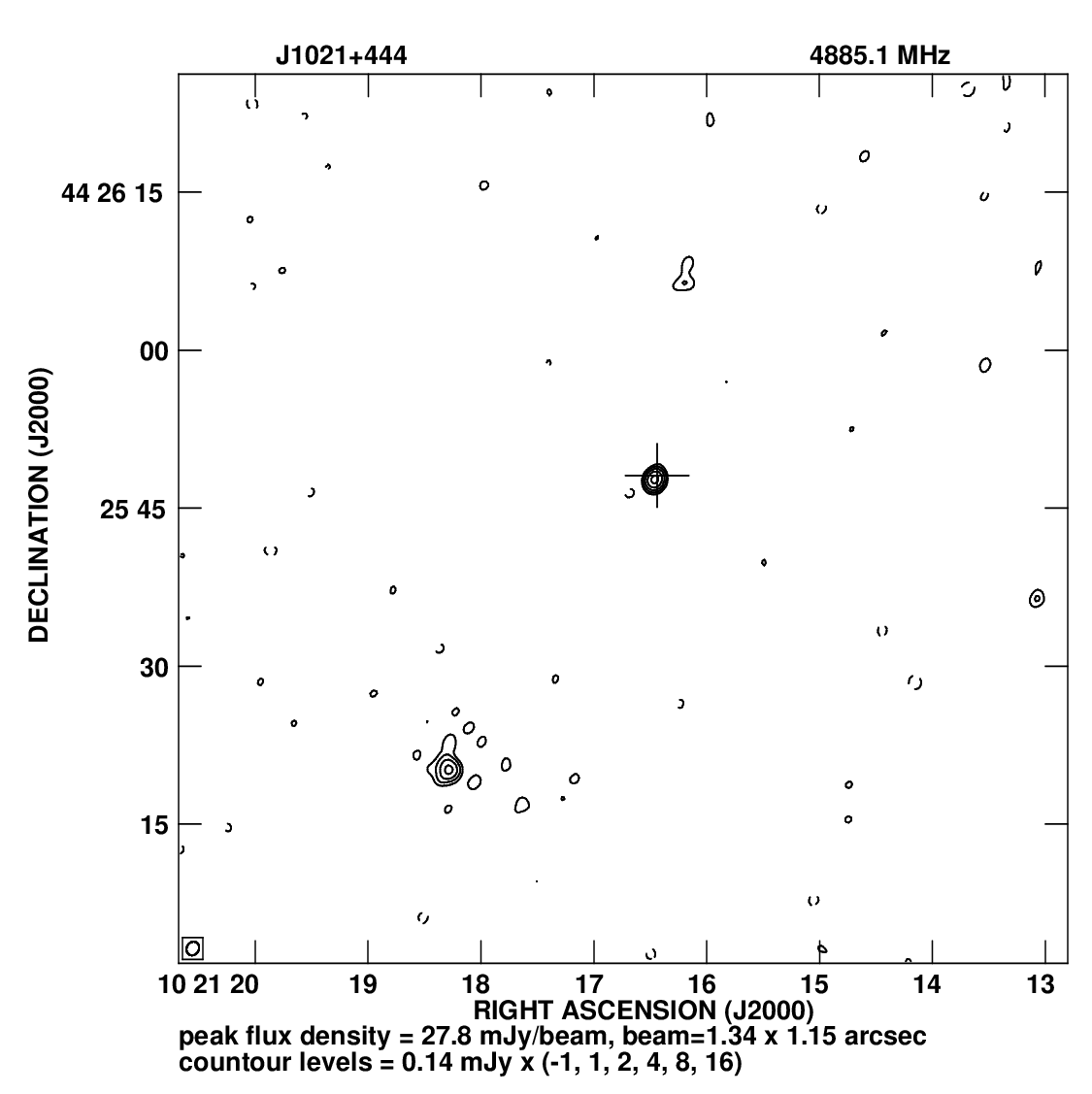}
{\small\\{\bf Fig.\,A.1.}
Sources rejected  as HYMORS candidates. \emph{Left panels}: FIRST
maps. \emph{Right panels}: 4.9\,GHz VLA B-conf. maps from this study.}
\end{figure*}

\begin{figure*}
\centering
\includegraphics[scale=0.4]{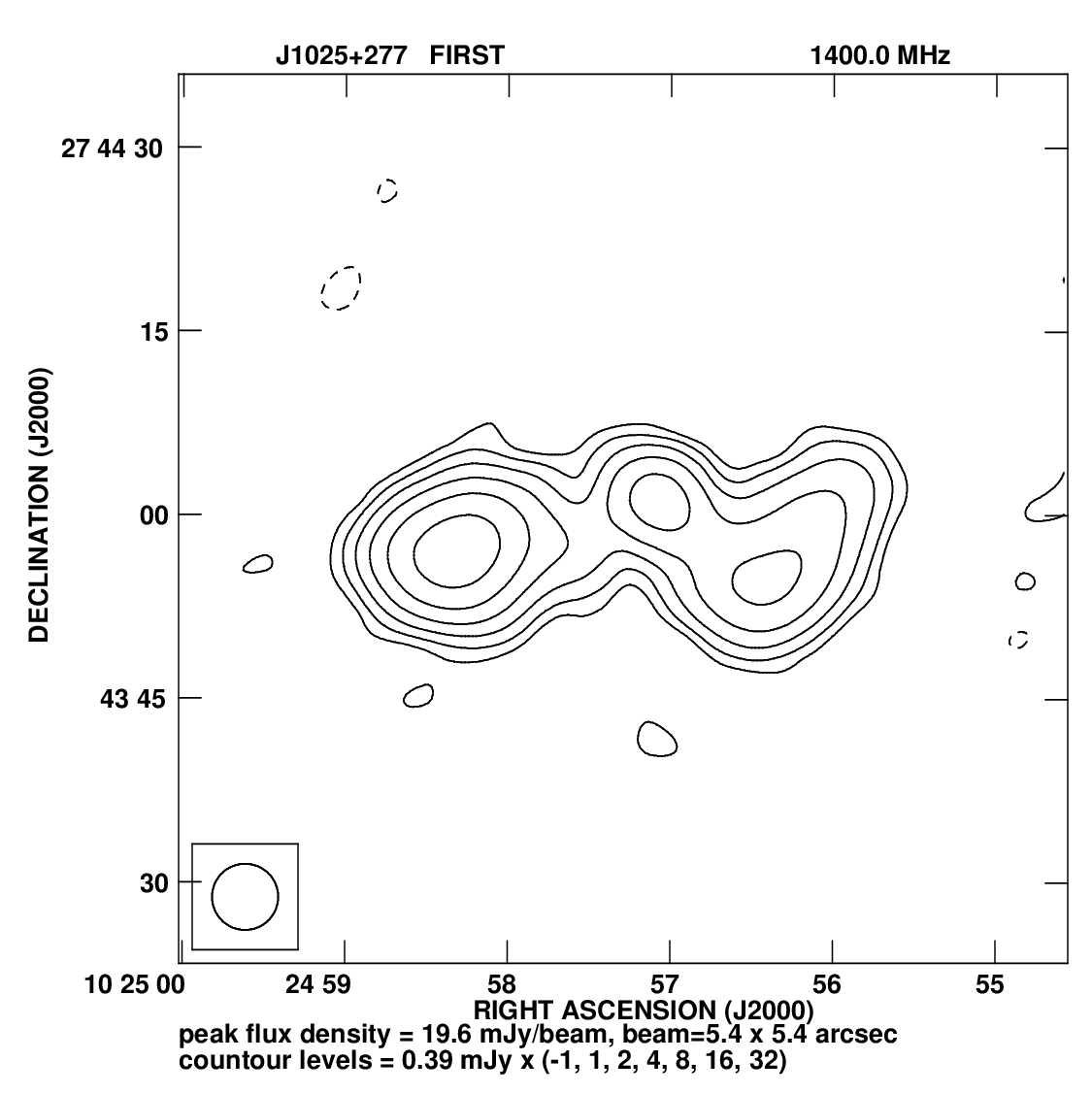}
\includegraphics[scale=0.4]{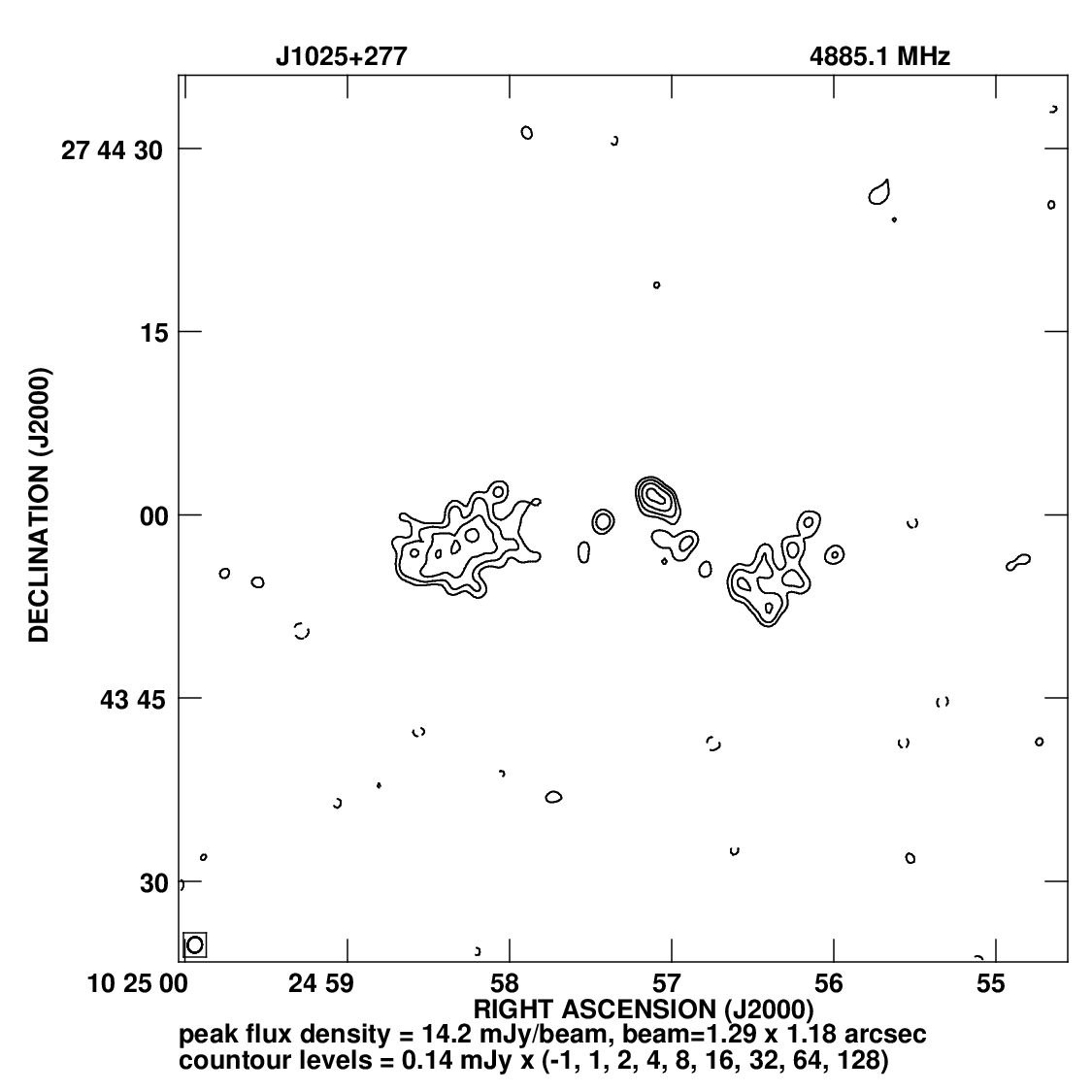}
\includegraphics[scale=0.4]{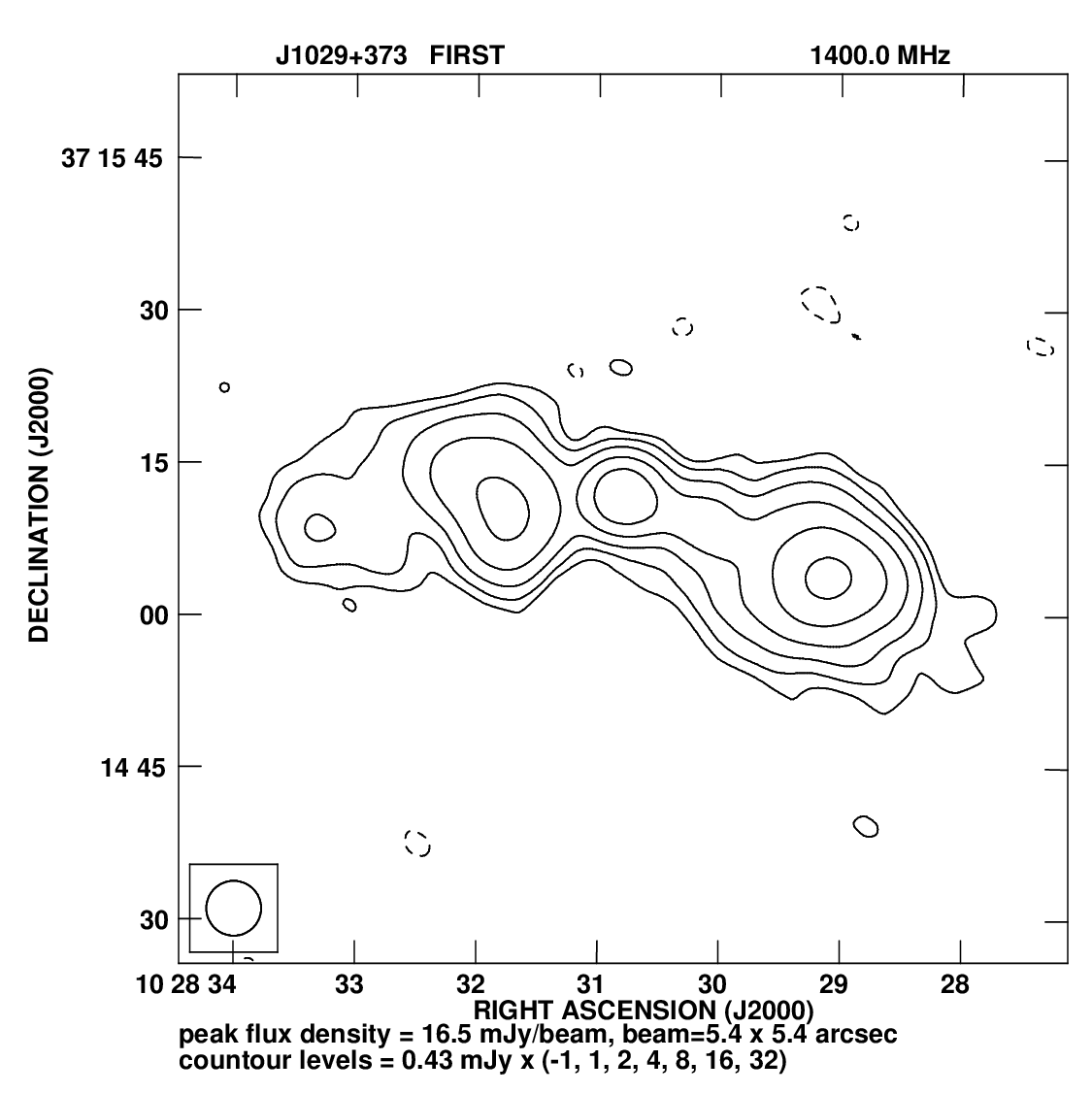}
\includegraphics[scale=0.4]{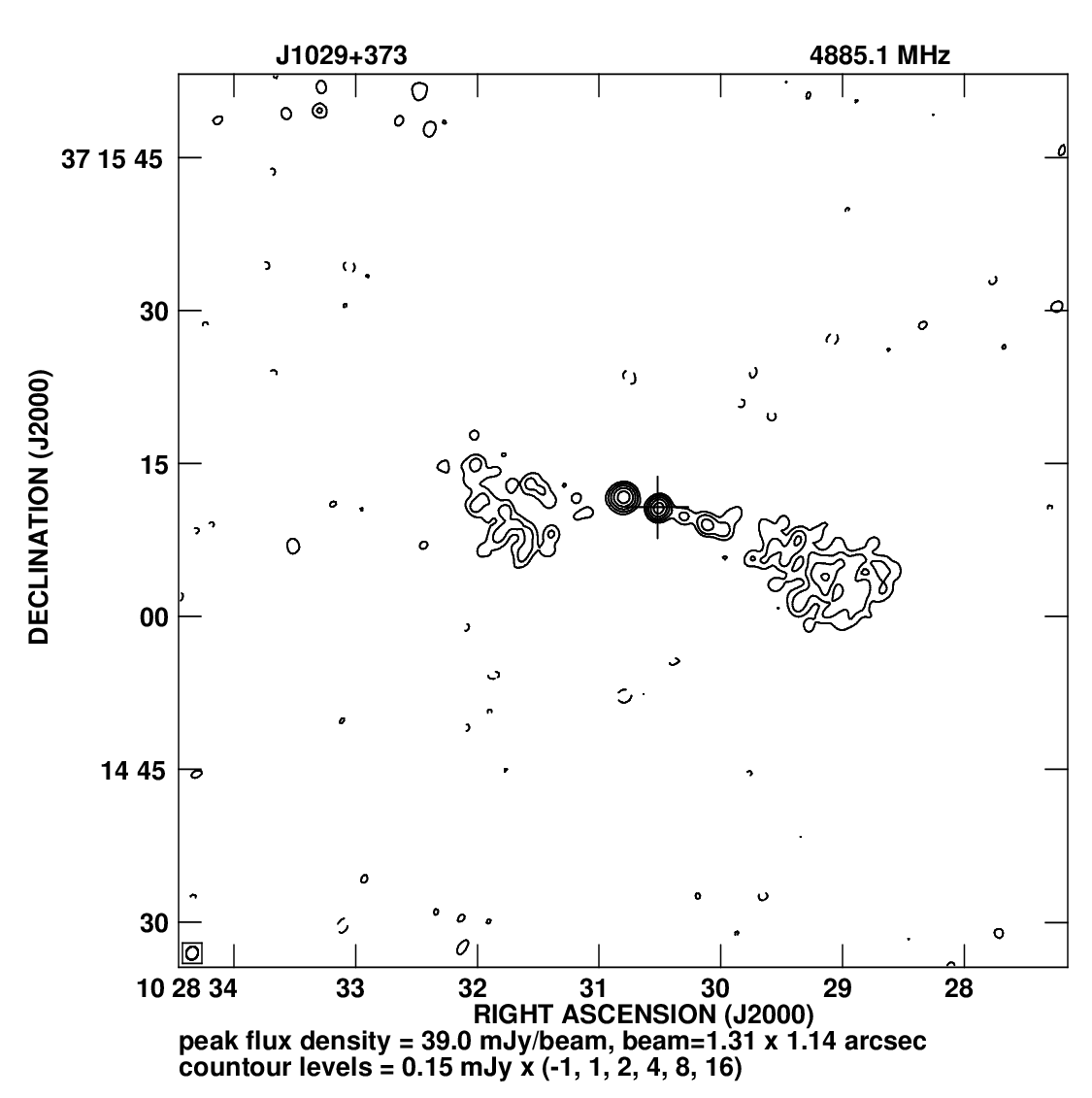}
\includegraphics[scale=0.4]{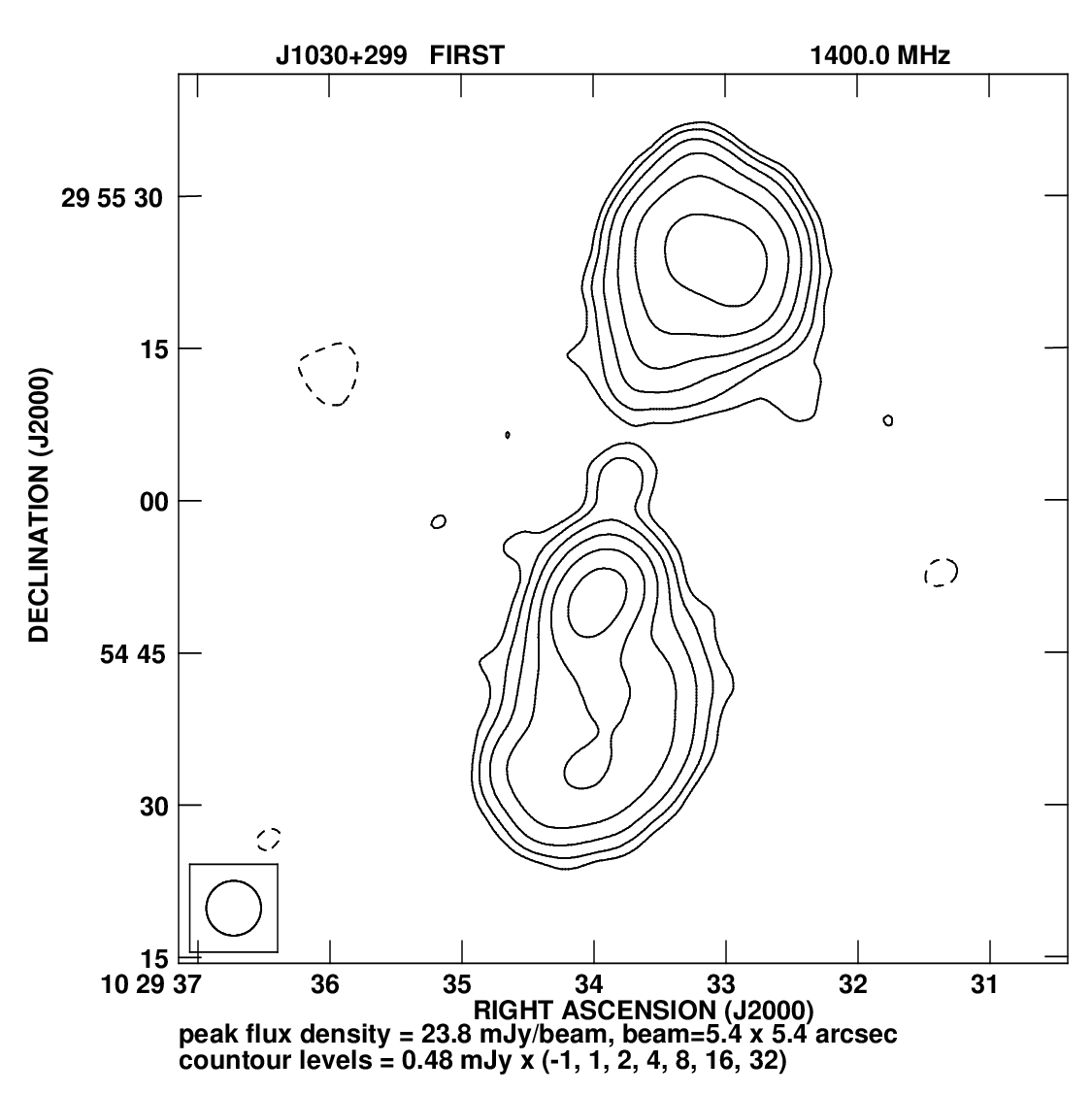}
\includegraphics[scale=0.4]{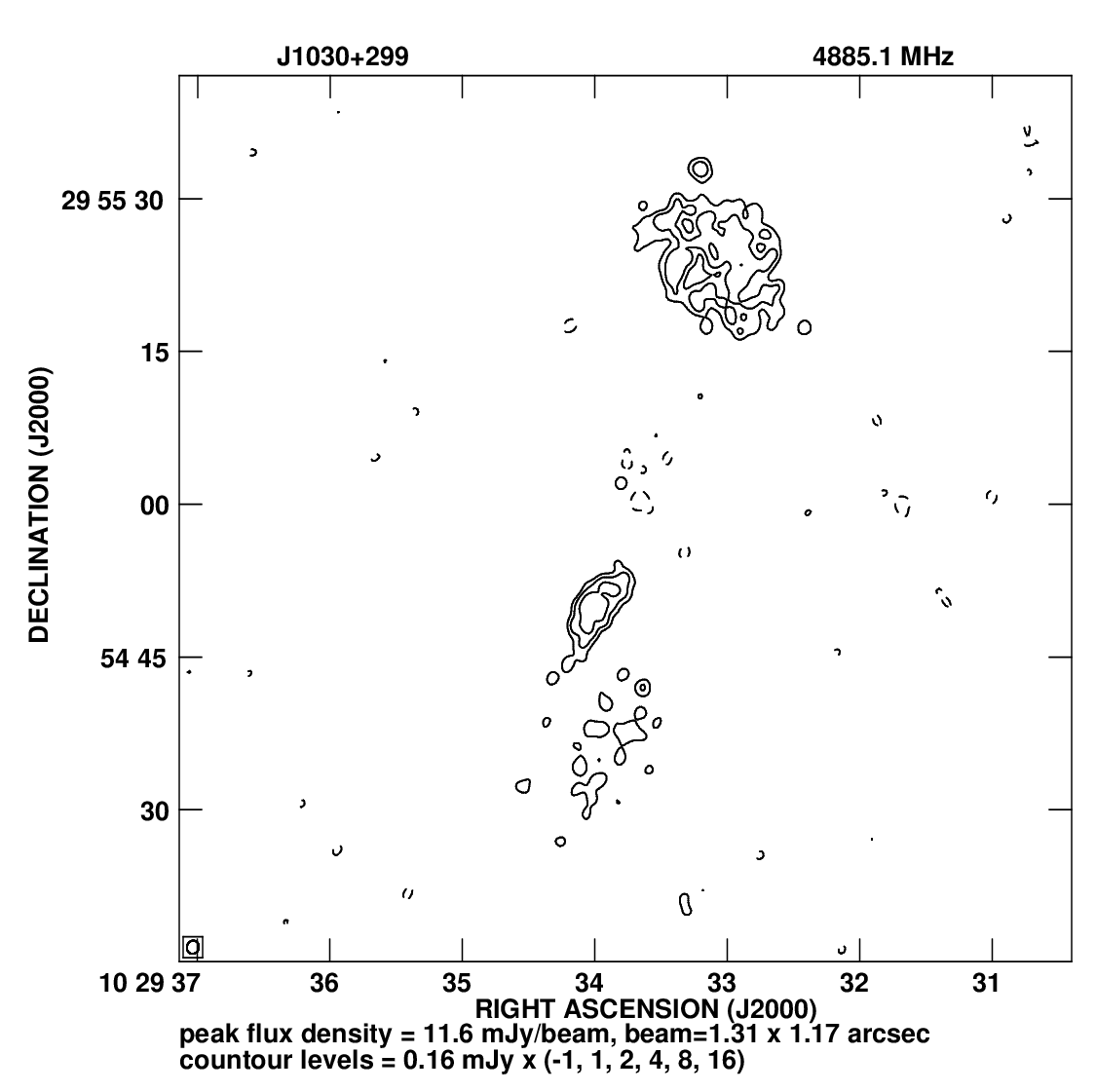}
{\small\\
{\bf Fig.\,A.1.}~(continued)}
\end{figure*}

\begin{figure*}
\centering
\includegraphics[scale=0.4]{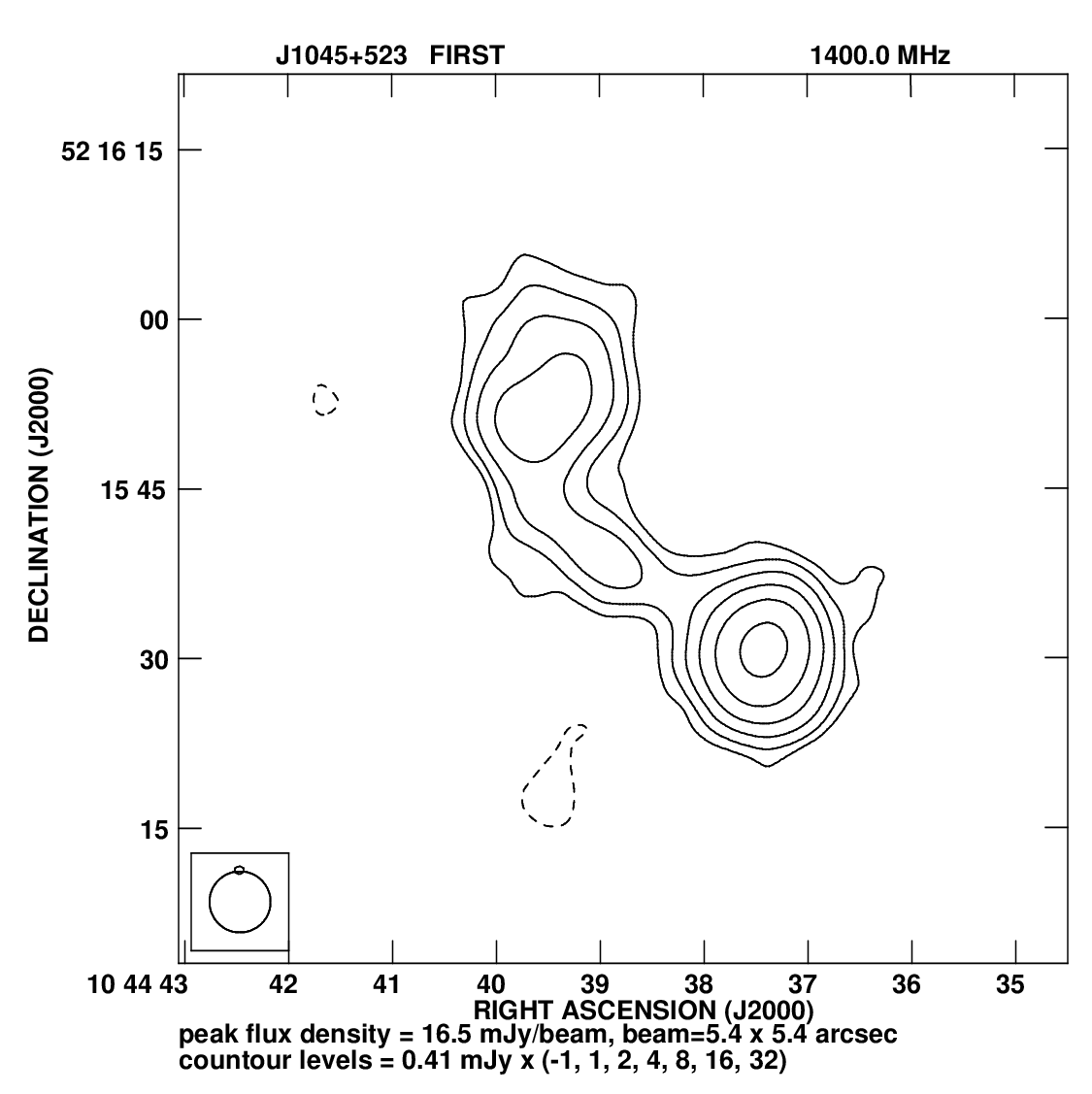}
\includegraphics[scale=0.4]{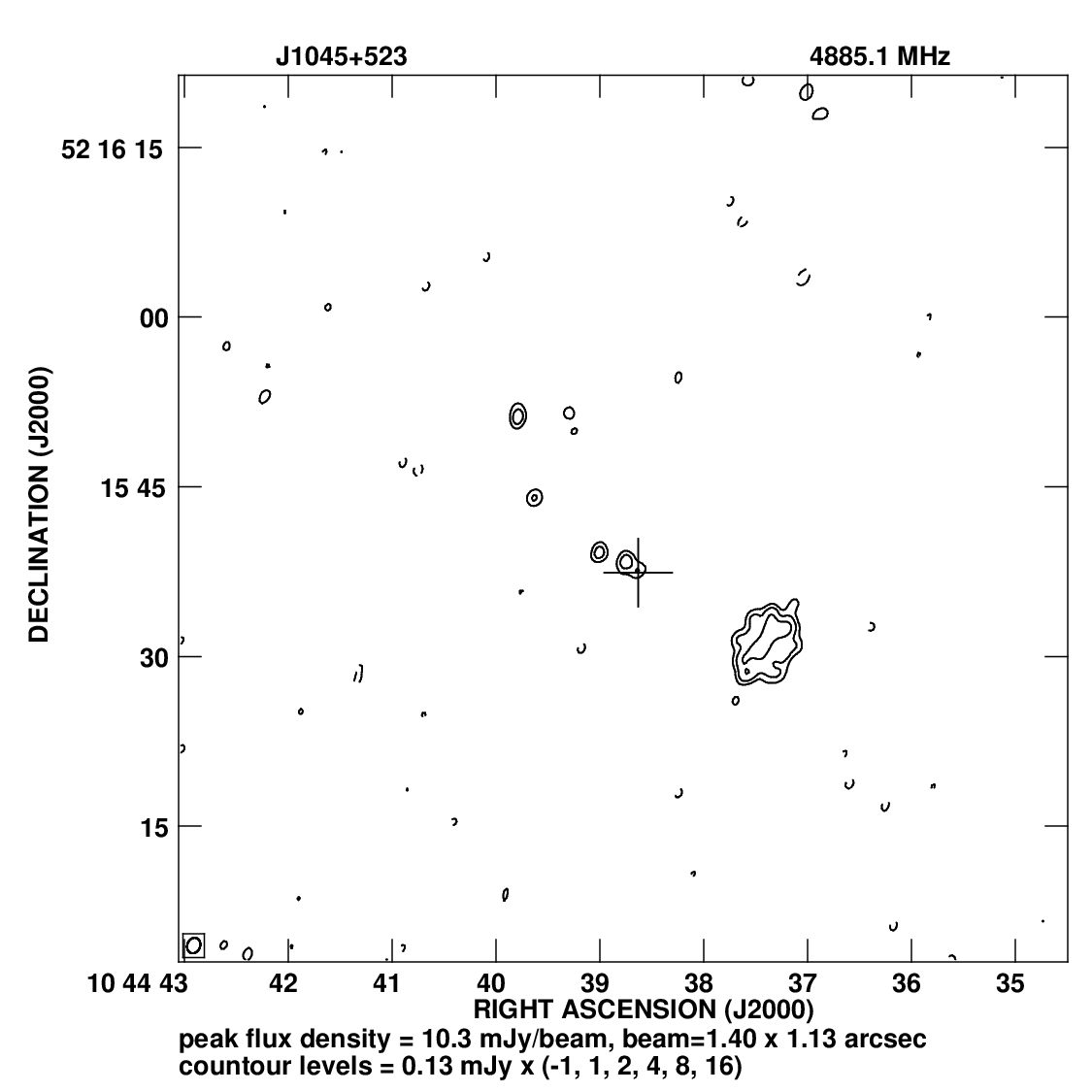}
\includegraphics[scale=0.4]{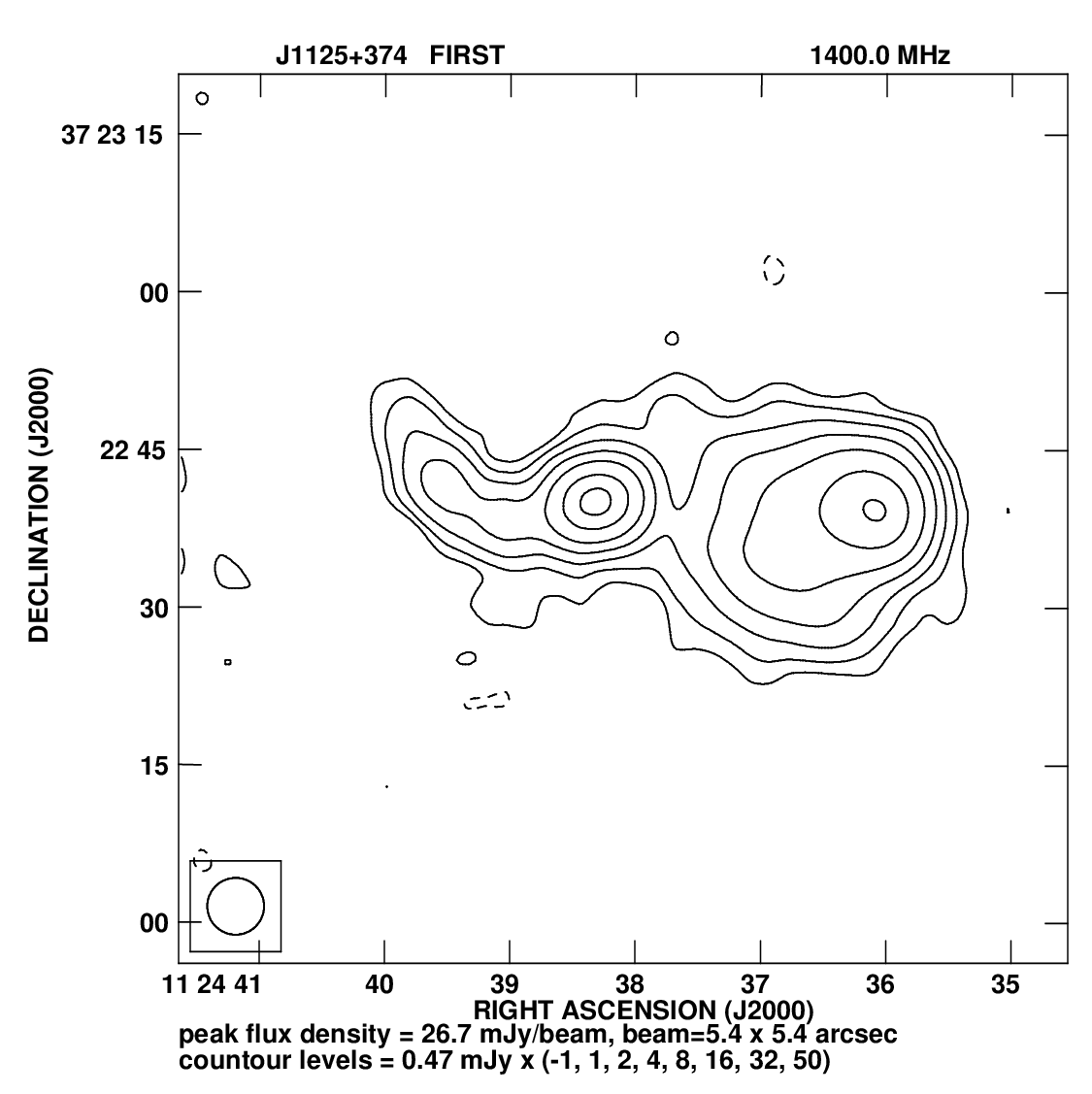}
\includegraphics[scale=0.4]{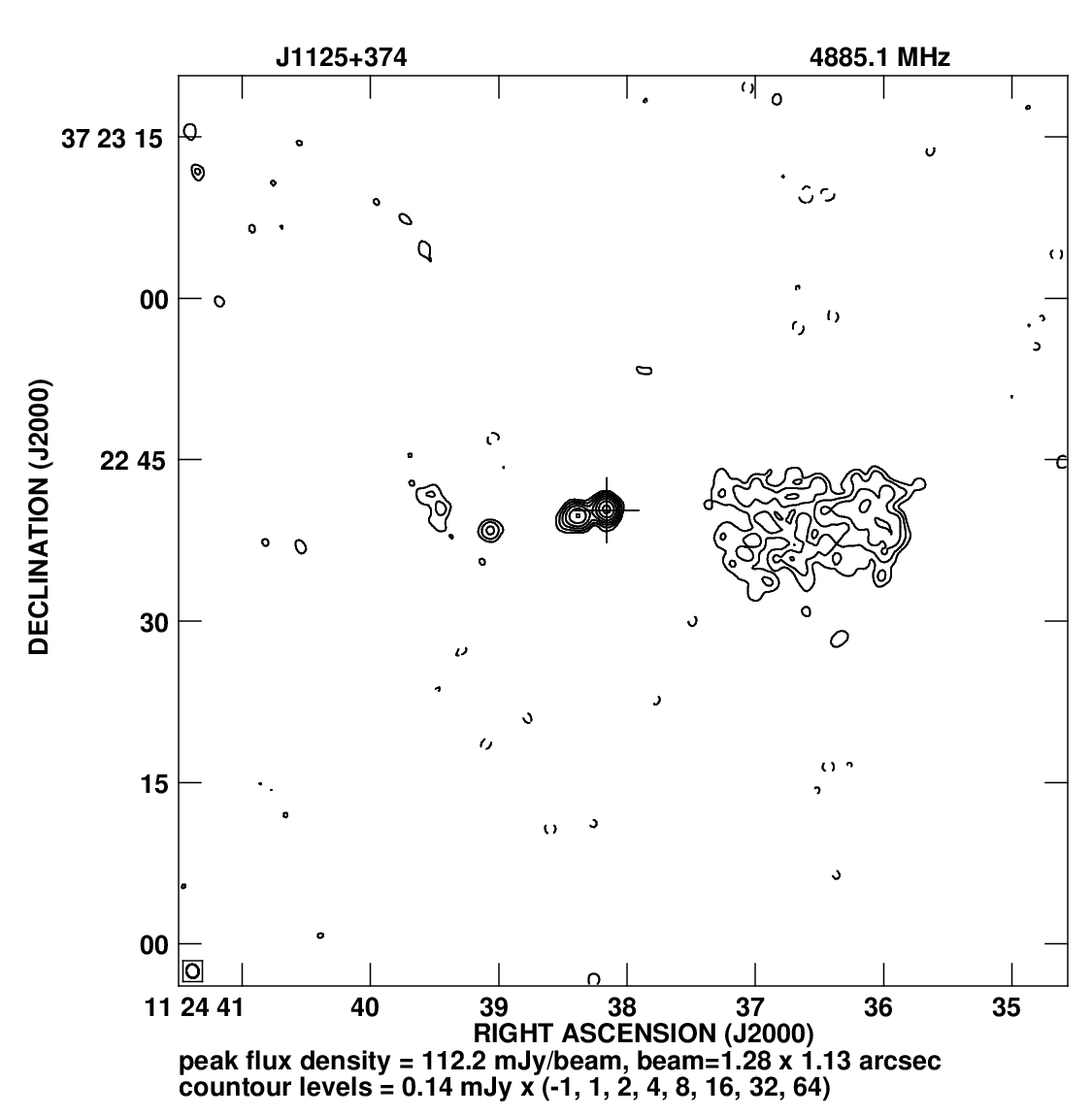}
\includegraphics[scale=0.4]{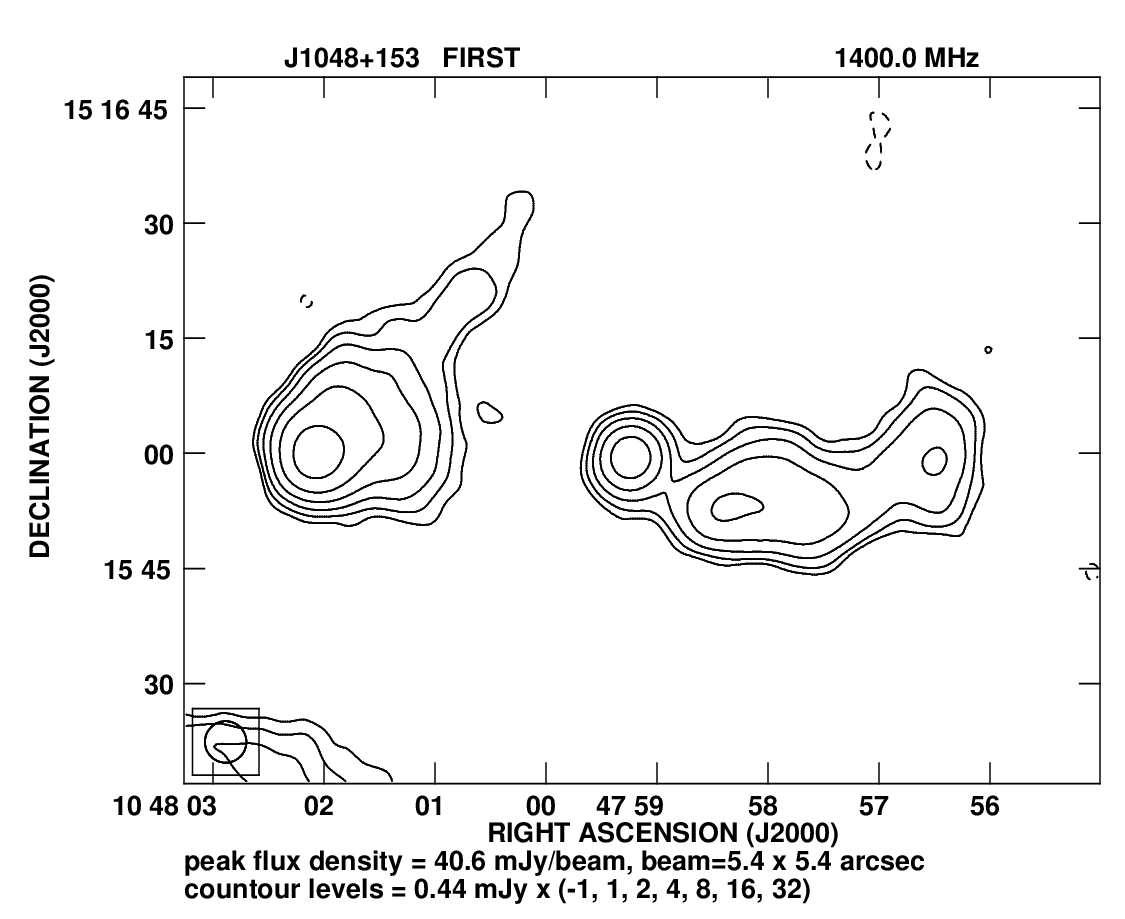}
{\small\\
{\bf Fig.\,A.1.}~(continued)}
\end{figure*}

\begin{figure*}
\centering
\includegraphics[scale=0.4]{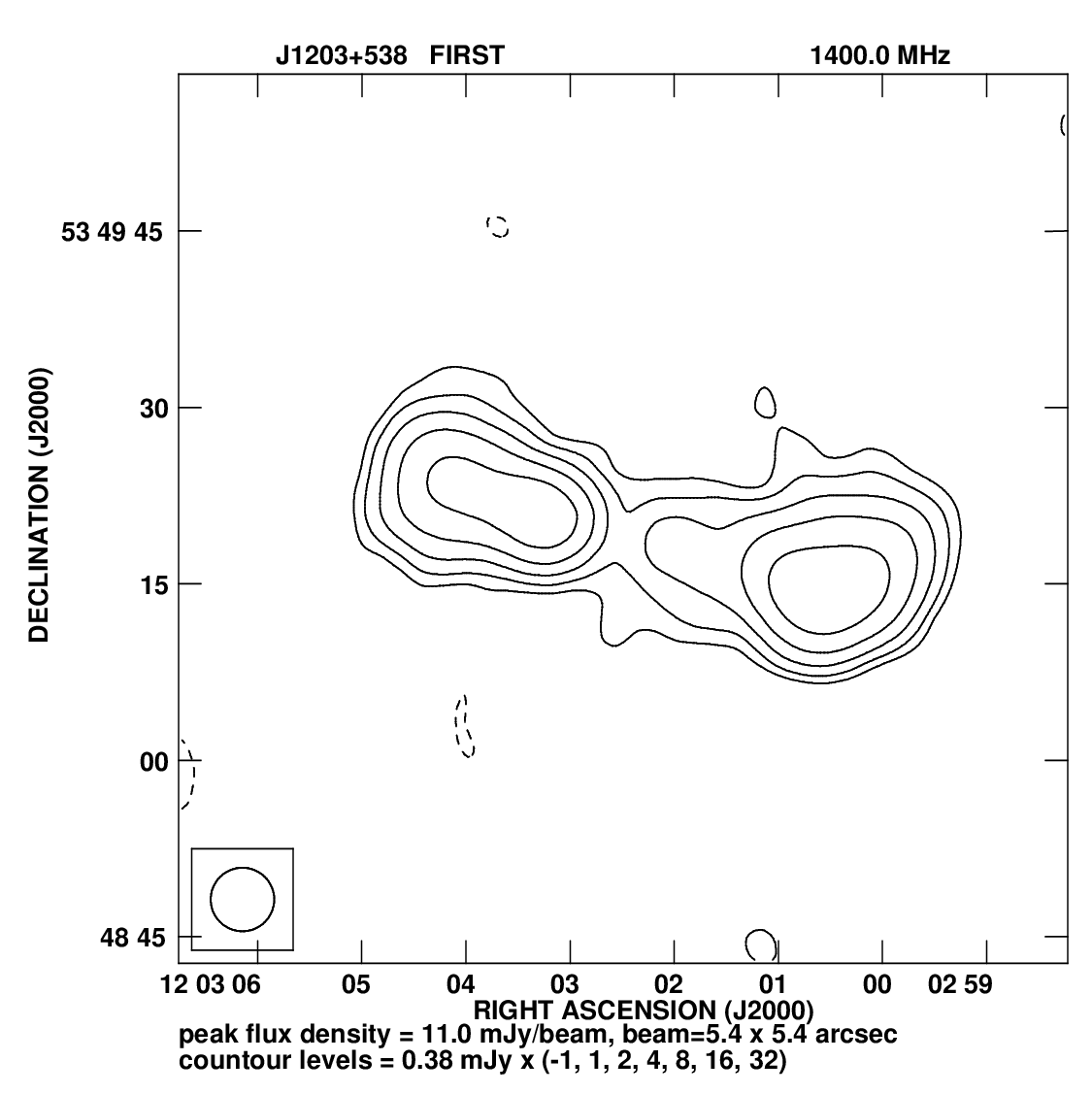}
\includegraphics[scale=0.4]{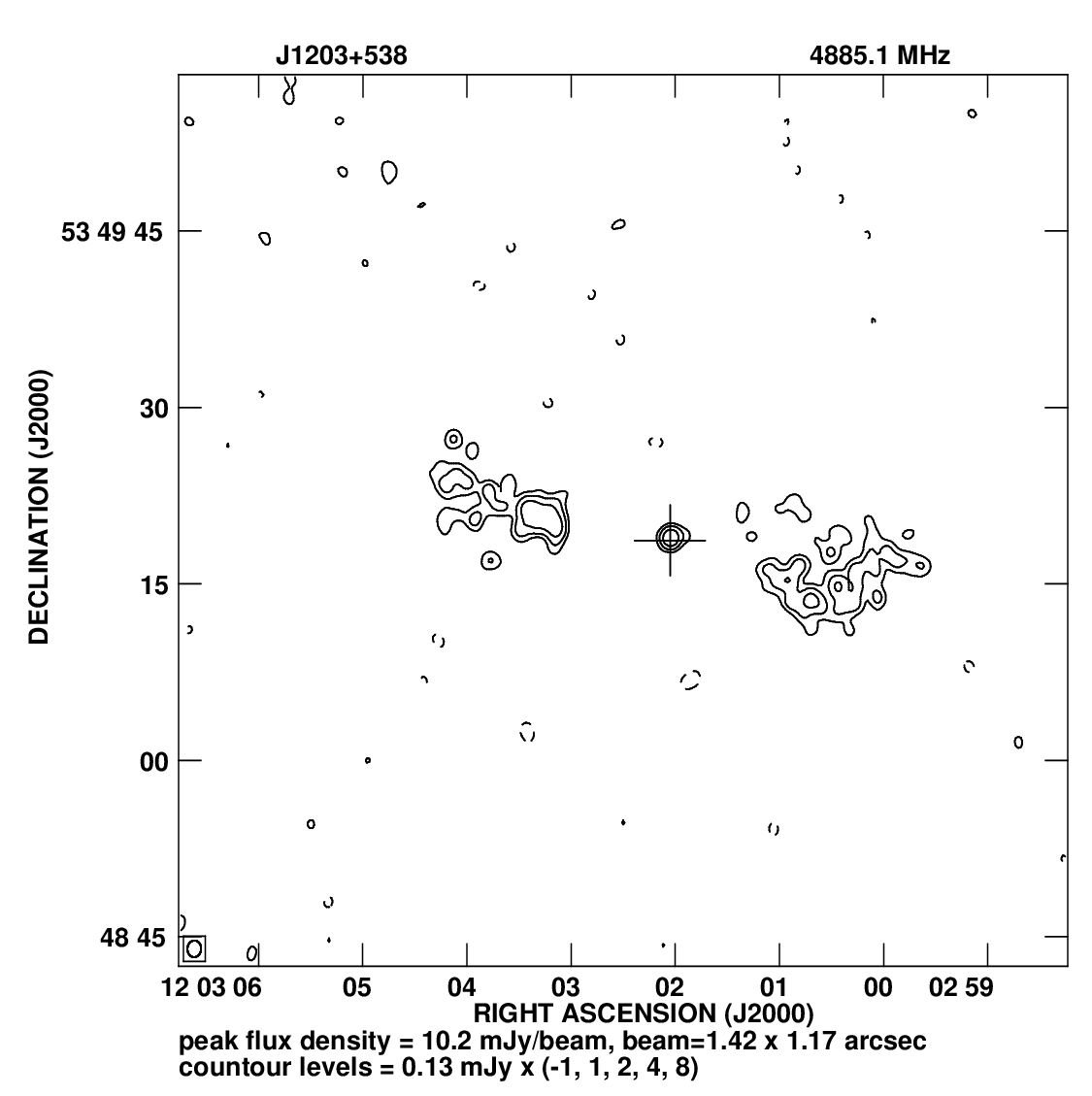}
\includegraphics[scale=0.4]{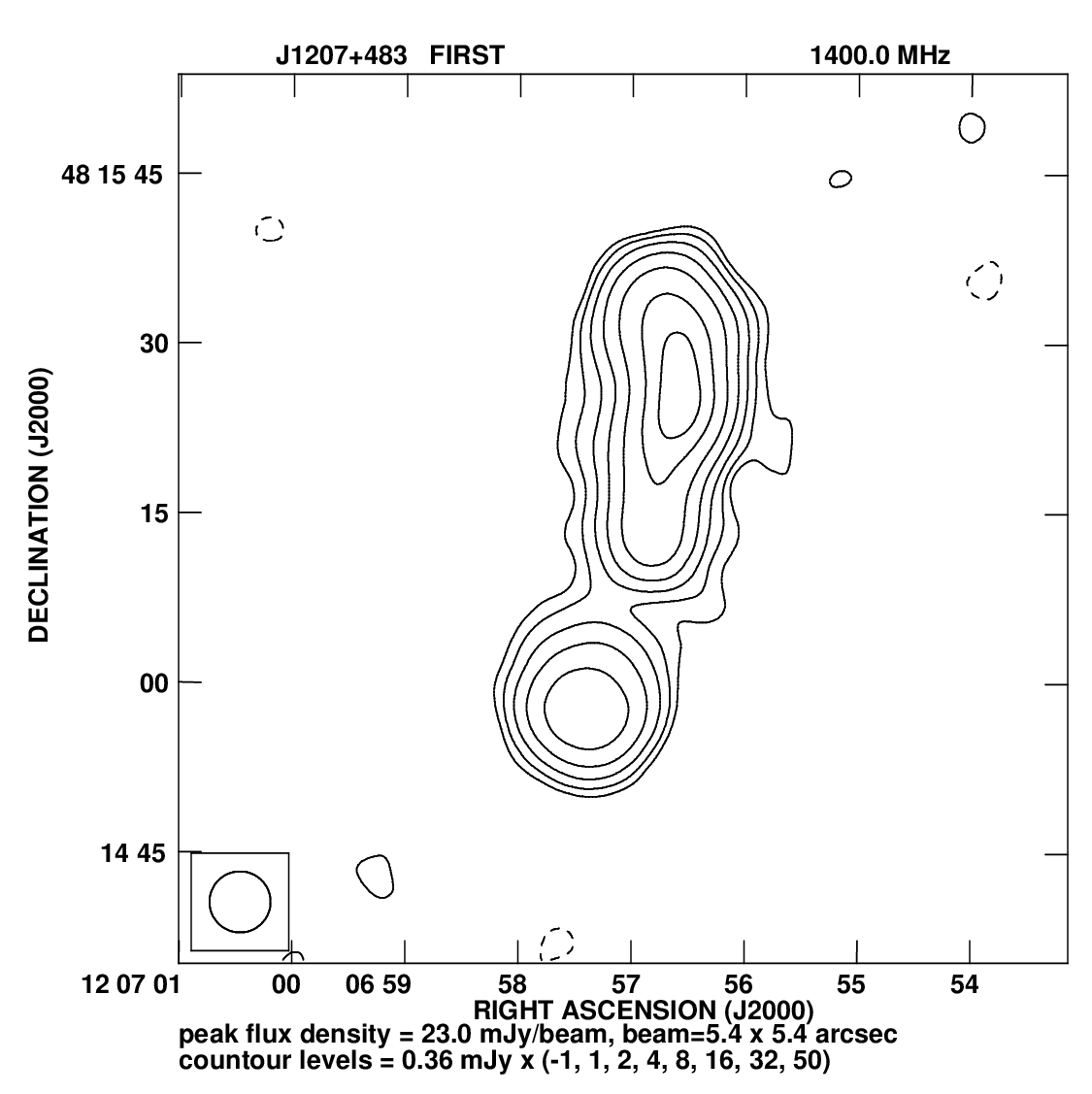}
\includegraphics[scale=0.4]{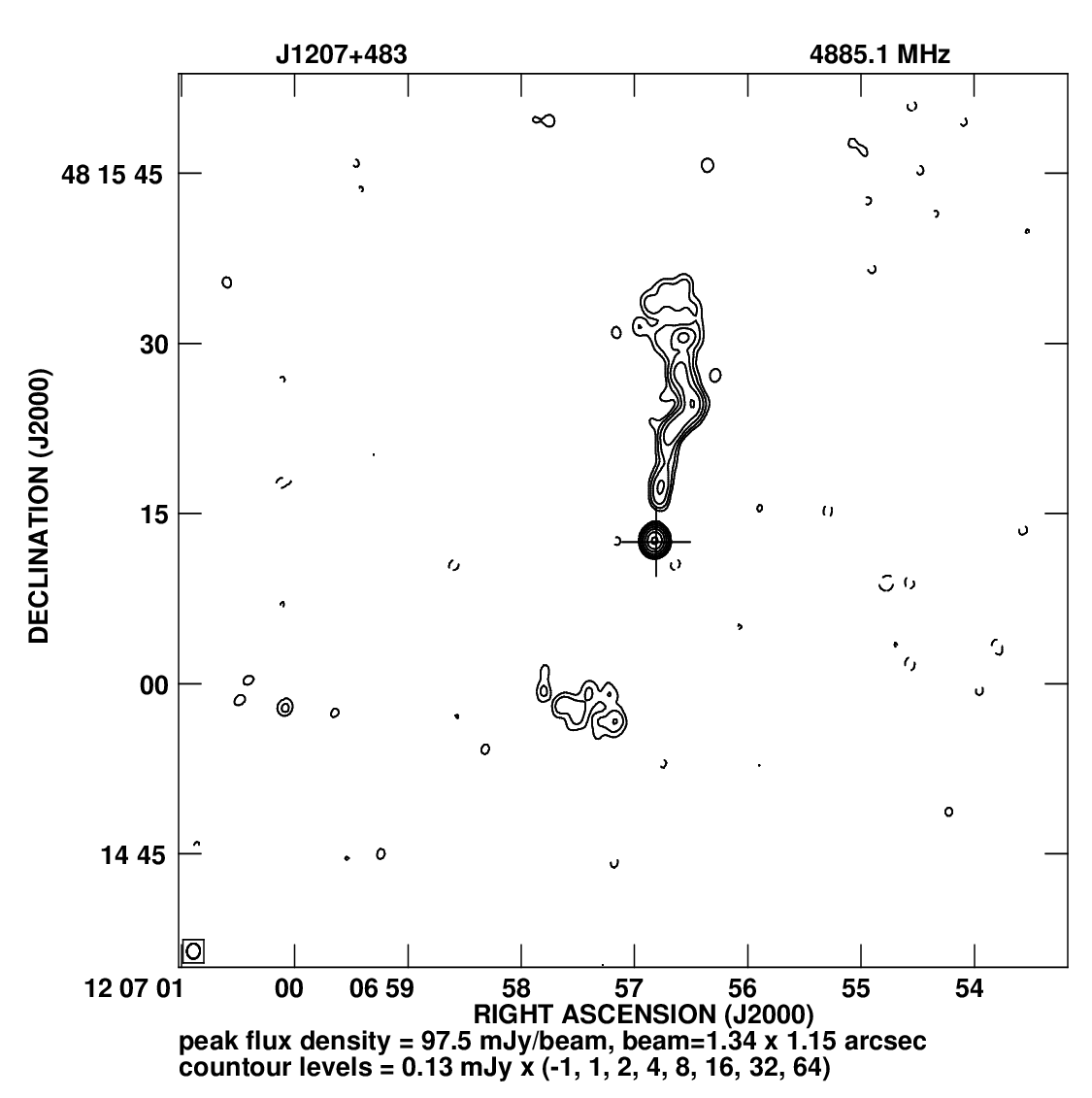}
\includegraphics[scale=0.4]{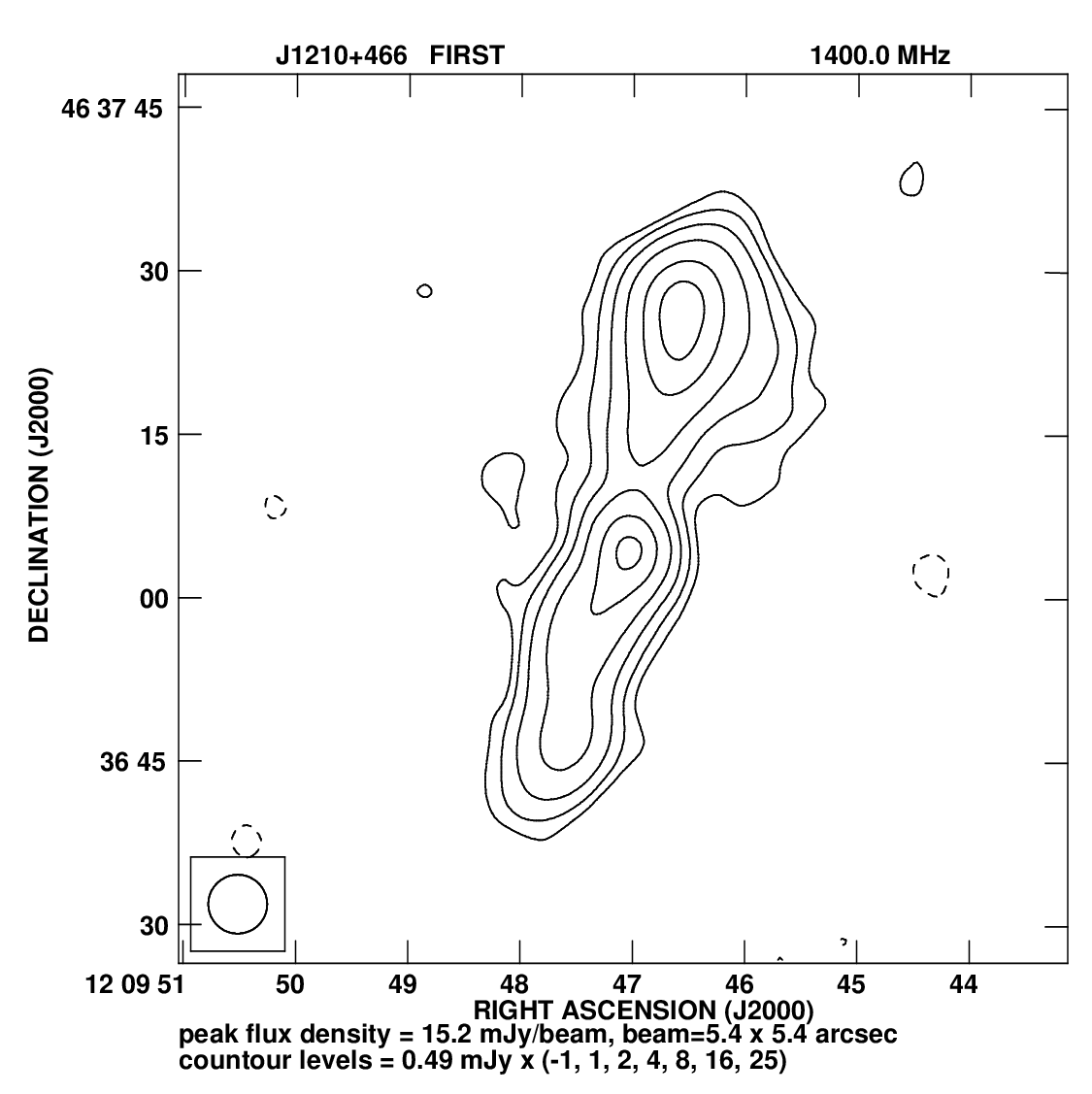}
\includegraphics[scale=0.4]{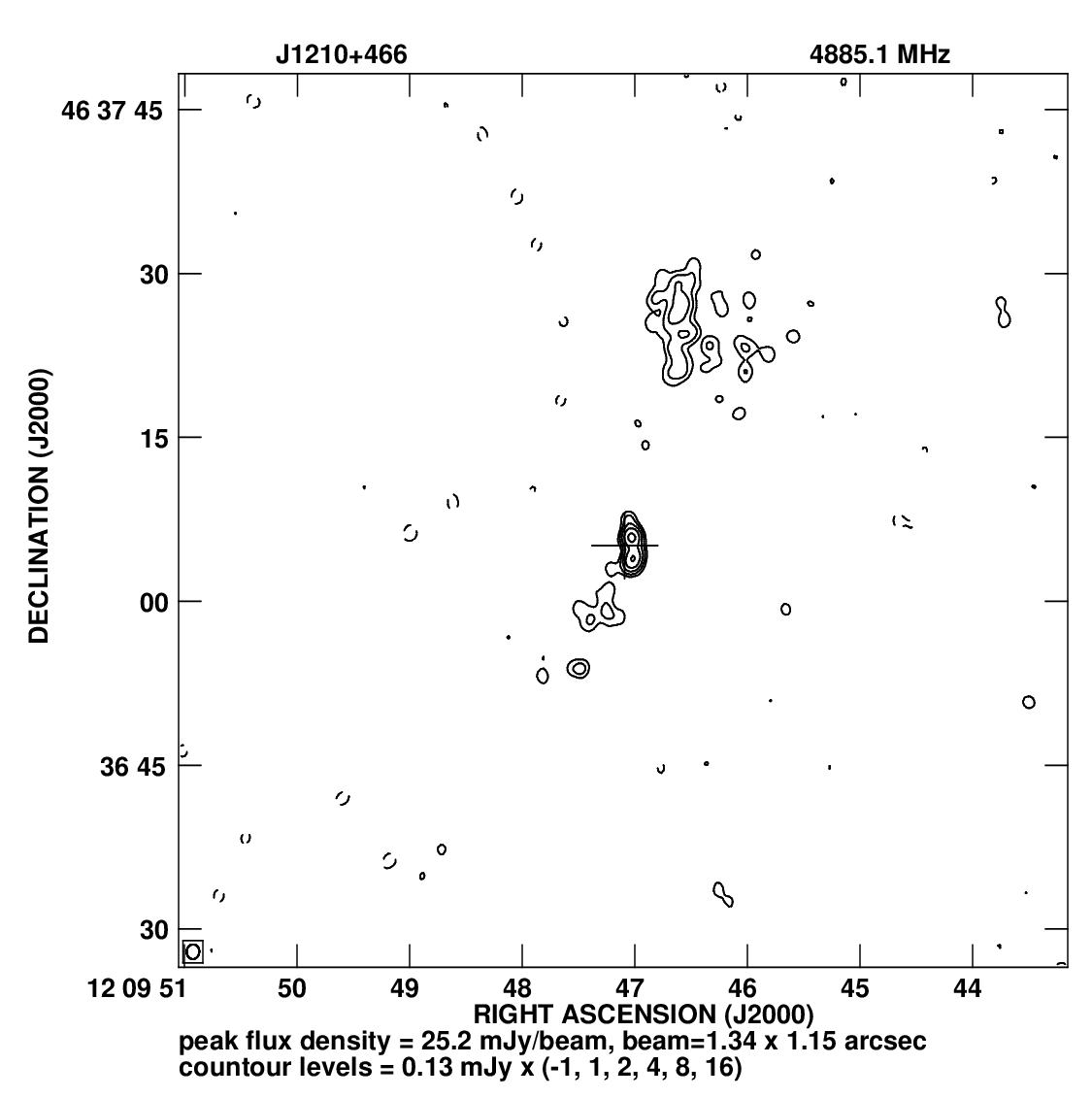}
{\small\\
{\bf Fig.\,A.1.}~(continued)}
\end{figure*}

\begin{figure*}
\centering
\includegraphics[scale=0.4]{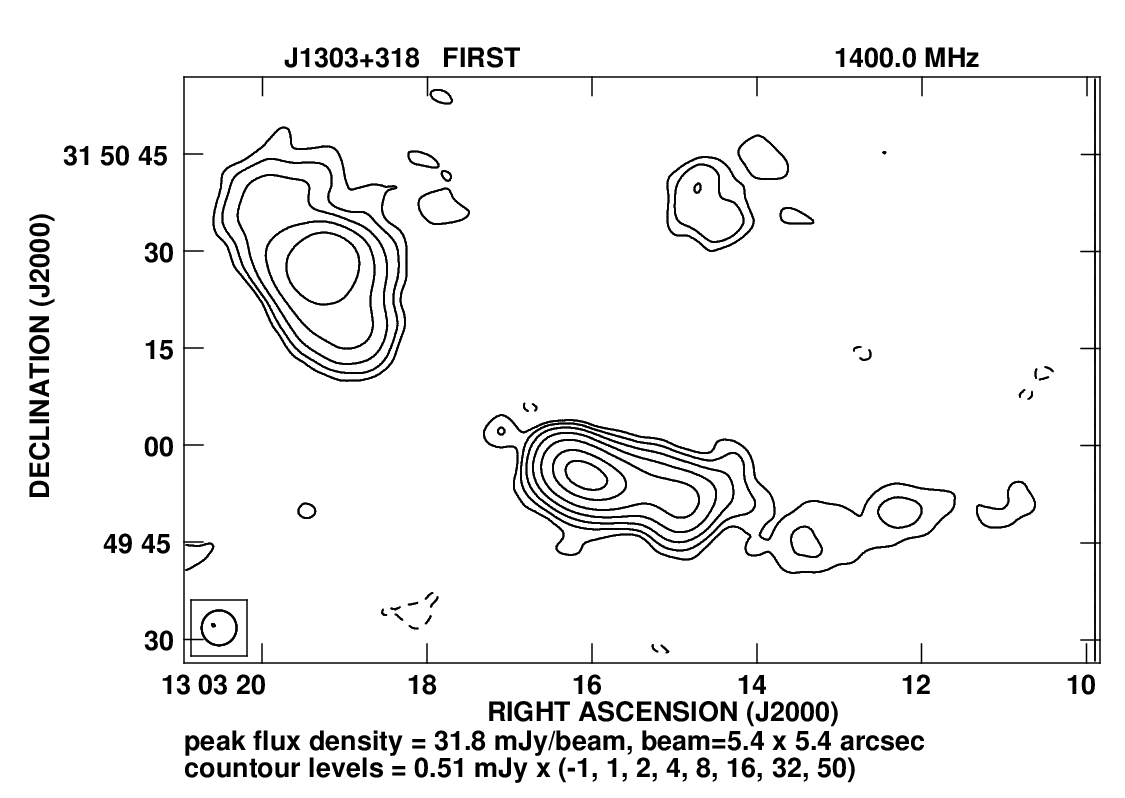}
\includegraphics[scale=0.4]{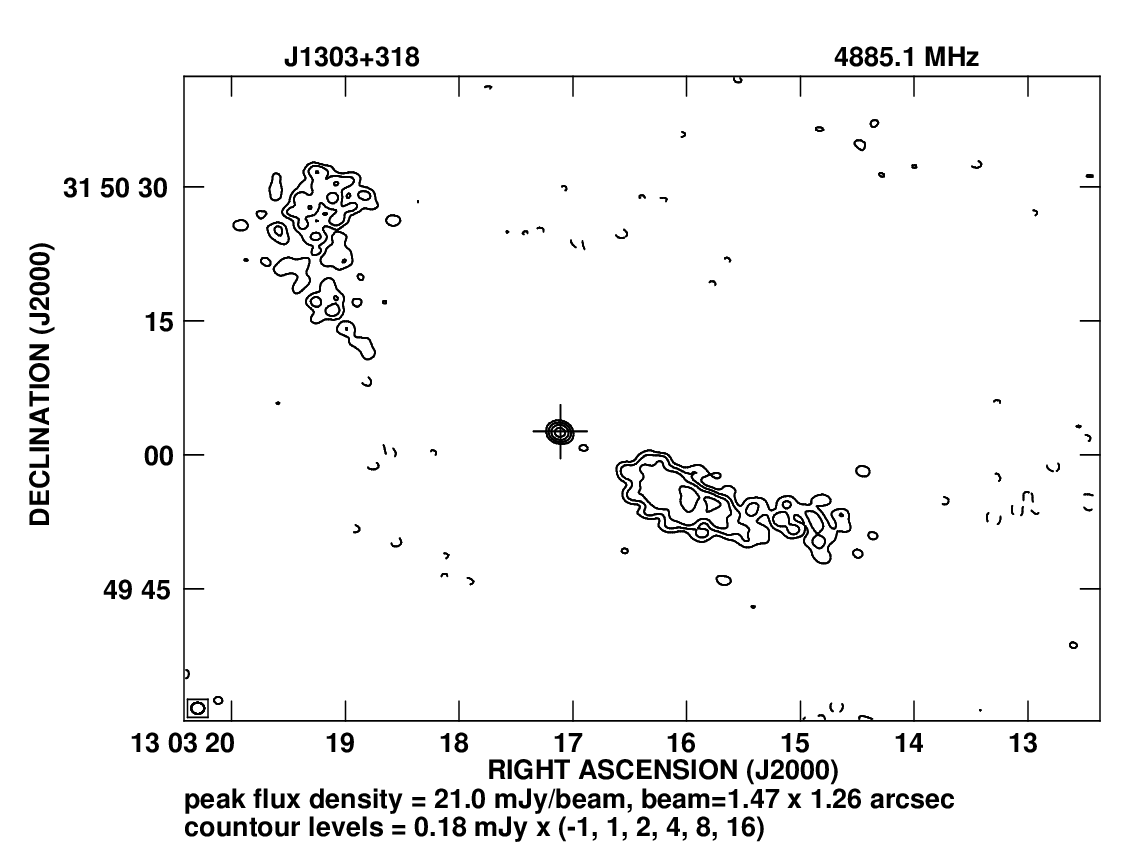}
\includegraphics[scale=0.4]{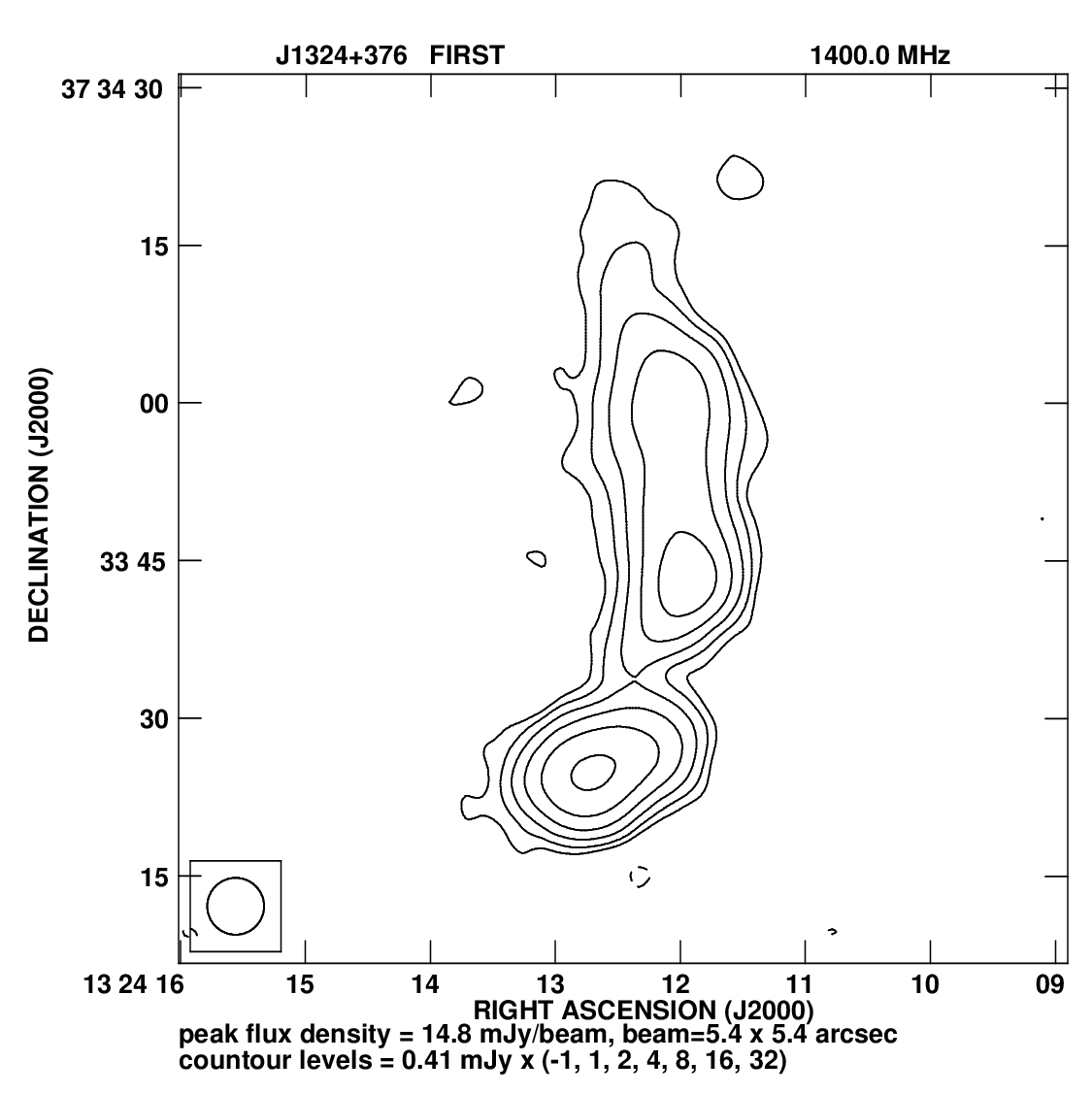}
\includegraphics[scale=0.4]{3996fA1A.ps}
\includegraphics[scale=0.4]{3996fA1B.ps}
\includegraphics[scale=0.4]{3996fA1C.ps}
{\small\\
{\bf Fig.\,A.1.}~(continued)}
\end{figure*}

\begin{figure*}
\centering
\includegraphics[scale=0.4]{3996fA1D.ps}
\includegraphics[scale=0.4]{3996fA1E.ps}
{\small\\
{\bf Fig.\,A.1.}~(continued)}
\end{figure*}

\end{document}